\documentclass[11pt,a4paper]{article}

\usepackage{titlesec}
\usepackage{amssymb}
\usepackage{amsmath,amsthm,amssymb,color}
\usepackage[pdftex,pagebackref,colorlinks]{hyperref}
\usepackage{graphicx}
\usepackage{multirow}
\usepackage{makecell}
\usepackage{booktabs}
\usepackage{enumerate}
\date{}
\usepackage{url}
\urldef{\mailsa}\path|{alfred.hofmann, ursula.barth, ingrid.haas, frank.holzwarth,|
\urldef{\mailsb}\path|anna.kramer, leonie.kunz, christine.reiss, nicole.sator,|
\urldef{\mailsc}\path|erika.siebert-cole, peter.strasser, lncs}@springer.com|

\newtheorem{theorem}{Theorem}
\newtheorem{lemma}{Lemma}
\newtheorem{corollary}{Corollary}
\newtheorem{definition}{Definition}
\newtheorem{proposition}{Proposition}

\newtheorem{remark}{Remark}
\newtheorem{example}{Example}

\newtheorem{construction}{Construction}

\begin{document}

\title{Minimal Binary Linear Codes from Vectorial Boolean Functions}

\author{Yanjun Li, Jie Peng\thanks{Corresponding author
\newline\indent Y. Li is with Institute of Statistics and Applied Mathematics, Anhui University of Finance {\&} Economics,  Bengbu, Anhui, 233030,   China,
(Email: yanjlmath90@163.com).
\newline \indent J. Peng is with Mathematics and Science College of
Shanghai Normal University, Guilin Road \#100, Shanghai, 200234, China. (Email:~jpeng@shnu.edu.cn).
\newline \indent H. Kan is with Shanghai Key Laboratory of Intelligent Information Processing, School of Computer Science, Fudan University; Shanghai Engineering Research Center of Blockchain, Shanghai Institute for Advanced Communication and Data Science, Shanghai 200433, China. (Email:~hbkan@fudan.edu.cn).
\newline \indent L. Zheng is  with the School of Mathematics and Physics, University of South China, Hengyang, Hunan, 421001, China. (Email:~zhenglijing817@163.com).}~,
   Haibin Kan,  and Lijing Zheng}
\maketitle
\begin{abstract}
 Recently, much progress has been made to construct minimal linear codes due to their preference in secret sharing schemes and secure two-party computation. In this paper, we put forward a new method to construct minimal linear codes by using vectorial Boolean functions. Firstly, we give a necessary and sufficient condition for a generic class of linear codes from vectorial Boolean functions to be minimal. Based on that, we derive some new three-weight minimal linear codes and determine their weight distributions. Secondly, we obtain a necessary and sufficient condition for another generic class of linear codes from vectorial Boolean functions to be minimal and to be violated the AB condition. As a result, we get three infinite families of minimal linear codes violating the AB condition. To the best of our knowledge, this is the first time that minimal liner codes are constructed from vectorial Boolean functions. Compared with other known ones, in general the minimal liner codes obtained in this paper have higher dimensions.
\end{abstract}



\section{Introduction}

Linear codes, especially those with few weights, paly an important role in consumer electronics, communication and data storage system. Recently, a special kind of linear codes, called minimal linear cods, has attracted great attention and research interest because of their significant applications in secret
sharing schemes \cite{Carlet-ding-2007-IT, Yuan-2006-IT} and secure two-party computation \cite{Chabanne-2014, Ding-2003}. They are those linear codes $C\subseteq \mathbb{F}_p^n$, in which every codeword is minimal, that is, every codeword $c\in C$ only covers $ac$ for any $a\in \mathbb{F}_p$, but no other codewords in $C$.

In 1998, a sufficient condition for judging whether a linear code is minimal was presented by Ashikhmin and Barg (see Lemma \ref{lemABcondition} and Remark \ref{characp} of this paper), which is called the AB condition. In the literature, several classes of minimal linear codes satisfying the AB condition are obtained by studying linear codes with few weights (see for instance \cite{Ding-2003, Ding-2015-IT, Hengzl-2016-FFA, Luogj-2018, Luogj-2018-2, Sihem-2017, Sihem-2019, Shimj-2017-IT, Tangcm-2016-IT, Wang-2015, Yangsd-2017-DCC, Zhouzc-2016-DCC}).
Although the AB condition is not necessary, there is not any infinite family of minimal linear codes violating the AB condition before 2018 until Chang and Hyun's break through in \cite{Chang-2018-DCC}, where the first infinite family of such codes was discovered amongst the linear code of the form
\begin{align}\label{eqcodeBooleanintro}
C_f=\bigg\{c(\mu,\nu)=\big(\mu f(x)+\nu \cdot x\big)_{x\in\mathbb{F}_{2}^{n*}} : \mu\in\mathbb{F}_2,\nu\in\mathbb{F}_{2}^{n}\bigg\},
\end{align}
where $f$ is a Boolean function on $\mathbb{F}_2^n$. Then a necessary and sufficient condition for the code $C_f$ to be minimal was given by Ding et al. in \cite{Ding-2018-IT}, and thus three new infinite families of minimal linear codes violating the AB condition were found. Later the same authors also extended their results to the case of finite fields of characteristic 3, and obtained an infinite family of linear codes with $\frac{w_{\rm min}}{w_{\rm max}}\leq \frac{2}{3}$ in \cite{Heng-2018-FFA}. Then Bartoli and Bonini \cite{Bartoli-2019-IT} further extended the results of \cite{Ding-2018-IT} and \cite{Heng-2018-FFA} to odd characteristic cases, and found an infinite family of minimal linear codes violating the AB condition. Thereafter,
more families of minimal linear codes violating the AB condition were constructed by investigating the linear code of the form \eqref{eqcodeBooleanintro}. For instance, some  minimal linear codes violating the AB condition were obtained from cutting blocking sets \cite{Bonini-2020-JAC, Pasalic-2021-CCDS};  from partial difference sets \cite{taoran-2021-IT}; from characteristic functions \cite{Sihem-2020-IT}; from  weakly regular plateaued/bent
functions \cite{Sihem-2020-IT2, Xu-2021-CCDS}; from Partial Spreads \cite{Xu-2019-IT}; from  Maiorana-McFarland functions \cite{Xu-2020-FFA, Zhangfengrong-2021-DCC}; and from the direct sum of Boolean functions \cite{Zhangfengrong-2022-DCC}, etc.

In the literature, there is another general form to construct minimal linear codes violating the AB condition, which is given by
\begin{align}\label{eqintro2}
C_D=\{(d_1\cdot x, d_2\cdot x,\ldots, d_s\cdot x) : x\in \mathbb{F}_p^n\},
\end{align}
where $p$ is a prime and $D:=\{d_1,d_2,\ldots, d_s\}\subseteq \mathbb{F}_p^n$ is called the definition set of $C_D$. Some conditions for $C_D$ to be minimal have been obtained in \cite{Luwen-2021,Tangcm-2021-IT}. Some related minimal linear codes violating the AB condition of this form can be found in \cite{Lixia-qinyue-2020, Luwen-2021, Tangcm-2021-IT, Zhangweiqin-2019-AAECC}.

This paper concentrates on the construction of minimal linear codes by using vectorial Boolean functions. Explicitly, we consider the minimality of binary linear codes of the form
\begin{align}\label{eqourconintro}
C_F=\bigg\{c(\mu,\nu)=\big(\mu\cdot F(x)+\nu\cdot x\big)_{x\in\mathbb{F}_{2}^{n*}} : \mu\in\mathbb{F}_2^m,\nu\in\mathbb{F}_2^n\bigg\},
\end{align}
where $F$ is any vectorial Boolean $(n,m)$-function. Note that some linear codes have been obtained from certain vectorial Boolean functions in the literature. For instance, the  parameters of the linear code
\begin{align}\label{eqintrocarletcon}
C_F=\bigg\{c(\mu,\nu,\lambda)=\big(\mu\cdot F(x)+\nu\cdot x+\lambda\big)_{x\in\mathbb{F}_{2}^{n}} : \mu\in\mathbb{F}_2^m,\nu\in\mathbb{F}_2^n,\lambda\in\mathbb{F}_2\bigg\}
\end{align}
were determined in \cite{Carlet-ding-2007-FFA} and \cite{Wadayama-2001}, where $F$ is an $(n,m)$-function. In \eqref{eqintro2}, let $D=\{x\in\mathbb{F}_{2^n}^* : {\rm Tr}_l^n\big(F(x)\big)=0\}$, the authors of \cite{Cheny-2020-FFA} obtained three classes of binary linear codes with few weights, and determined their weight distributions, where $F$ is a power function on $\mathbb{F}_{2^n}$. Modifying the linear code in \eqref{eqintro2} as
\begin{align*}
C_{D_{f_\lambda}}=\bigg\{\big({\rm Tr}_1^m(xd_1)+{\rm Tr}_1^s(yF(d_1)),\ldots, {\rm Tr}_1^m(xd_{n_{f_\lambda}})+{\rm Tr}_1^s(yF(d_{n_{f_\lambda}})) \big) : x\in\mathbb{F}_{2^m},y\in\mathbb{F}_{2^s}\bigg\},
\end{align*}
the authors of \cite{Tangdeng-2017} obtained several classes of linear codes with few weights and determine their weight distributions by choosing some suitable $(m,s)$-functions $F$, where $\lambda\in\mathbb{F}_{2^s}^*$, $f_\lambda(x)={\rm Tr}_1^s(\lambda F(x))$, $n_{f_\lambda}$ is the Hamming weight of $f_\lambda$ and $D_{f_\lambda}$ is  the support of $f_\lambda$. In \cite{Ding-2019-IT}, Ding et al. used vectorial bent functions for a construction of a two-parameter family of binary linear codes that do not satisfy the conditions of the Assmus-Mattson theorem. However, it is harder to construct minimal linear codes of the form \eqref{eqourconintro} than that of the form \eqref{eqcodeBooleanintro}, since if $C_F$ in \eqref{eqourconintro} is minimal, then for any nonzero Boolean function $f=\mu\cdot F$, the linear code $C_f$ in \eqref{eqcodeBooleanintro} is always minimal. To the best of our knowledge, there has not been any infinite families of minimal linear codes constructed from vectorial Boolean functions up to now.

In this paper, we first give a necessary and sufficient condition such that the linear code $C_F$ in \eqref{eqourconintro} is minimal. Based on that, we obtain some three-weight minimal linear codes from vectorial plateaued functions and determine their weight distributions.
 Then, by rewriting the $(n, m)$-function $F$ as $F=(f, G)$ (where $f$ is a Boolean function on $\mathbb{F}_2^n$ and $G$ is an $(n, m-1)$-function), we give another necessary and sufficient condition such that $C_F$ in \eqref{eqourconintro} is minimal and violates the AB condition.
 This condition enables us to derive three infinite families of minimal linear codes violating the AB condition.
Moreover, we find that the dimensions of our minimal linear codes violating the AB condition are also better than almost all known ones. The parameters of known minimal linear codes over $\mathbb{F}_2$ violating the AB condition are listed in Table \ref{intro}, from which the advantage of dimensions of our codes can be seen clearly.

\begin{table*}[!t]\setlength{\abovecaptionskip}{0cm}
 \small
\centering
\caption{Known minimal linear codes over $\mathbb{F}_2$ violating the AB condition}\label{intro}
\medskip
\begin{tabular}{|c|m{53pt}<{\centering}|m{53pt}<{\centering}|c|c|c|}
 \hline
Form & Length & Dimension & Minimal distance & Method & References\\
 \hline
  &  $2^n-1$  &$n+1$ & $2^{|A|}+2^{|B|}-2^{|A\cap B|}-1$&  &\cite[Prop 3]{Chang-2018-DCC}  \\
\cline{2-4} \cline{6-6}
 & $2^n-1$   & $n+1$ & $s(2^t-1)$, $n=2t$, $s\leq 2^{t-2}$ &  & \cite[Thm 18]{Ding-2018-IT} \\
 \cline{2-4} \cline{6-6}
 & $2^n-1$   & $n+1$ & $2^{n-1}-2^{\frac{n-1}{2}-1}(\frac{n+1}{2}-1)$ &  & \cite[Thm 23]{Ding-2018-IT} \\
 \cline{2-4} \cline{6-6}
 & $2^n-1$   & $n+1$ & $\sum_{j=1}^{k} {n\choose j}$, $2\leq k\leq \lfloor\frac{n-3}{2}\rfloor$ &  & \cite[Thm 31]{Ding-2018-IT} \\
  \cline{2-4} \cline{6-6}
 & $2^n-1$   & $n+1$ & $2^{t-1}(s+1)$, $n=s+t$ &  & \cite[Thm 4.1]{Xu-2020-FFA} \\
 \cline{2-4} \cline{6-6}
  & $2^n-1$   & $n+1$ & $2^{t-1}(2^k-1)d$, $n=2k+t$ & Boolean & \cite[Thm 4.2]{Xu-2020-FFA} \\
  \cline{2-4} \cline{6-6}
  & $2^n$   & $n+1$ & $2^{n-1}-2^{\lambda-1}(\kappa-1)$, $n=\lambda+\kappa$ & functions & \cite[Thm 2(i)]{Zhangfengrong-2021-DCC} \\
  \cline{2-4} \cline{6-6}
  & $2^n$   & $n+1$ & $2^{n-1}-2^{\lambda-1}(\kappa-3)$, $n=\lambda+\kappa$ &  & \cite[Thm 2(ii)]{Zhangfengrong-2021-DCC} \\
  \cline{2-4} \cline{6-6}
\eqref{eqcodeBooleanintro} & $2^n$   & $n+1$ & $2^{n-1}-2^{\lambda-1}(2^\kappa-\kappa-3)$ &  & \cite[Thm 3]{Zhangfengrong-2021-DCC} \\
   \cline{2-4} \cline{6-6}
  & $2^n$   & $n+1$ & $2^{n-1}-2^{\lambda}(\kappa+1)$ &  & \cite[Thm 4]{Zhangfengrong-2021-DCC} \\
   \cline{2-4} \cline{6-6}
  & $2^n$   & $n+2$ & $2^{n-1}-2^{\lambda}(\kappa+1)$ &  & \cite[Thm 5]{Zhangfengrong-2021-DCC} \\
  \cline{2-4} \cline{6-6}
  & $2^n$   & $n+\frac{s}{2}+2$ & $2^{\frac{r-1}{2}+s}(\frac{r+1}{2}+1)$, $n=r+s$ &  & \cite[Thm 7]{Zhangfengrong-2021-DCC} \\
   \cline{2-4} \cline{6-6}
  & $2^n-1$   & $n+1$ & $2^{\frac{n-1}{2}-1}(\frac{n+1}{2}+1)-1$ &  & \cite[Thm 3.5]{Sihem-2020-IT} \\
    \cline{2-4} \cline{6-6}
  & $2^n-1$   & $n+1$ & $2^{\frac{n-1}{2}-1}(\frac{n+1}{2}+1)-1$ &  & \cite[Thm 3.5]{Sihem-2020-IT} \\
    \cline{2-4} \cline{6-6}
  & $2^n-1$   & $n+1$ & - &   & \cite{Sihem-2020-IT} \\
   \cline{2-6}
  & $2^n-1$   & $n+1$ & - & Blocking  & \cite{Bonini-2020-JAC}  \\
   \cline{2-4} \cline{6-6}
  & $2^n$   & $n+1$ & $ |\Delta|$  &  sets  & \cite{Pasalic-2021-CCDS}  \\
    \cline{2-6}
  & $2^n-1$   & $n+1$ & $2^{n-1}+\theta_2$  & P-D. sets  & \cite{taoran-2021-IT}  \\
   \hline
    & $N_1$  & $n$ &  $d_1$ &   & \cite[Thm 3.7]{Lixia-qinyue-2020}  \\
    \cline{2-4} \cline{6-6}
    & $N_2$   & $n$ & $d_2$  & Boolean & \cite[Thm 3.14]{Lixia-qinyue-2020}  \\
      \cline{2-4} \cline{6-6}
    &$N_3$   & $n$ & $d_3$ & functions & \cite[Thm 3.20]{Lixia-qinyue-2020}  \\
    \cline{2-4} \cline{6-6}
    & $N_4$    & $n$ & $d_4$ &  & \cite[Thm 3.26]{Lixia-qinyue-2020}  \\
     \cline{2-6}
      \eqref{eqintro2}  & ${n\choose 2}$   & $n-1$ & $n-1$ &   & \cite[Thm 3]{Zhangweiqin-2019-AAECC}   \\
         \cline{2-4} \cline{6-6}
        & ${n\choose 2}+1$   & $n-1$ & $n$ &  - & \cite[Thm 4]{Zhangweiqin-2019-AAECC}   \\
         \cline{2-4} \cline{6-6}
        & ${n\choose 1}+{n\choose 2}$   & $n$ & $n$ &   & \cite[Thm 5]{Zhangweiqin-2019-AAECC}   \\
         \cline{2-4} \cline{6-6}
        & ${n\choose 1}+{n\choose 2}+1$   & $n$ & $n+1$ &   & \cite[Thm 6]{Zhangweiqin-2019-AAECC}   \\
         \hline
         & $2^n-1$   & $n+2$ & $2^{\frac{n}{2}}-2$ &   & Thm \ref{thfirstvioab}   \\
         \cline{2-4} \cline{6-6}
   \makecell{ \\ \eqref{eqourconintro} }   & $2^n-1$   & $n+m~(2<m\leq \frac{n}{2}+1)$ & $2^{n-2}+2^{\frac{n}{2}+1}$ & \makecell{Vectorial \\ Boolean \\ functions} & Thm \ref{th-2<m<t}   \\
       \cline{2-4} \cline{6-6}
      & $2^n-1$   & $2n+1$  & $2^{n-2}+2^{\frac{n}{2}+1}$ &   & Thm \ref{thGold}   \\
         \hline
\end{tabular}
  {the values of $N_i$ and $d_i$, $i=1,2,3,4$, are referred to \cite{Lixia-qinyue-2020}}\hspace{6cm}
\end{table*}

\medskip

The rest of this paper is organized as follows. In Section \ref{sec:Preliminaries}, we fix some notations used in this paper, and introduce some basic knowledge on Boolean functions and linear codes. In Section \ref{sec:generic1}, we define a generic construction of linear codes from vectorial Boolean functions, and fully characterize its minimality and parameters. Based on that, we give some 3-weight minimal linear codes with complete weight distribution in Section \ref{sec:threeweight}. In Section \ref{sec:generic2}, we present a necessary and sufficient condition for a generic class of linear codes to be minimal and to be violated the AB condition. Based on that, we construct several minimal linear codes violating the AB condition in Section \ref{sec:violating}. Finally, we conclude the paper in Section \ref{sec:conclusion}.

\section{Preliminaries}\label{sec:Preliminaries}

 Throughout this paper, $n$ and $m$ are two positive integers. For any set $E$, let $E^{*}=E \setminus\{0\}$ and let $|E|$ be the cardinality of $E$.
 Let $\mathbb{F}_{2^n}$ be the finite field of $2^n$ elements, and $\mathbb{F}_2^n$ be the $n$-dimensional linear space over $\mathbb{F}_2$. When fix a basis $\{\alpha_1,\alpha_2,\ldots,\alpha_n\}$ of $\mathbb{F}_{2^n}$ over $\mathbb{F}_2$, there is a natural one-to-one correspondence between $\mathbb{F}_{2^n}$ and $\mathbb{F}_2^n$, that is, for every $a\in\mathbb{F}_{2^n}$, there is a unique vector $(a_1,a_2,\ldots,a_n)\in\mathbb{F}_2^n$, such that $a=a_1\alpha_1+a_2\alpha_2+\cdots+a_n\alpha_n$. Thus in this paper, $\mathbb{F}_{2^n}$ and $\mathbb{F}_2^n$ are often treated as the same.

   The \emph{trace function} ${\rm Tr}_r^n : \mathbb{F}_{2^n}\rightarrow\mathbb{F}_{2^r}$, where $r|n$, is defined as
\begin{align*}
{\rm Tr}_r^n(x)\!=\!x\!+\!x^{2^r}\!+\!x^{2^{2r}}\!+\!\cdots\!+\!x^{2^{n-r}}, \hspace{0.2cm} \forall~x\in\mathbb{F}_{2^n}.
\end{align*}
For $r=1$, ${\rm Tr}_1^n$ is also called the \emph{absolute trace function}. It is well known that the trace function satisfies ${\rm Tr}_r^n(x)={\rm Tr}_r^n(x^{2^r})$, ${\rm Tr}_r^n(x+y)={\rm Tr}_r^n(x)+{\rm Tr}_r^n(y)$ ($\mathbb{F}_{2^r}$-linear) and ${\rm Tr}_r^n(x)={\rm Tr}_r^k({\rm Tr}_k^n(x))$ (transitivity property) for any $r| k$ and $k| n$.

\subsection{Vectorial Boolean function and Walsh-Hadamard transform}

A mapping  $F$ from $\mathbb{F}_2^n$ to $\mathbb{F}_2^m$ is called an {\it $(n,m)$-function}, and $F$ is called a {\it Boolean function} on $\mathbb{F}_2^n$ if $m=1$. For an $(n,m)$-function $F$ and a vector $\mu\in \mathbb{F}_2^{m*}$, the Boolean function $\mu\cdot F(x)$ is said to be the {\it component} of $F$ with respect to $\mu$. The {\it Walsh-Hadamard transform} of $\mu\cdot F$ at $\nu\in\mathbb{F}_2^n$ is defined as
\begin{align*}
W_{\mu\cdot F(x)}(\nu)=\sum_{x\in\mathbb{F}_2^n}(-1)^{\mu\cdot F(x)+\nu\cdot x},
\end{align*}
 where $\nu\cdot x$ is the usual product of $\nu$ and $x$ in $\mathbb{F}_2^n$. In $\mathbb{F}_{2^n}$, one has $\nu\cdot x={\rm Tr}_1^n(\nu x)$. For convenience, $W_{\mu\cdot F(x)}(\nu)$ is also denoted by $W_F(\mu,\nu)$ in this paper. The {\it nonlinearity} of $F$ is defined as
\begin{align*}
NL(F)=2^{n-1}-\frac{1}{2}\max_{\mu\in\mathbb{F}_2^{m*},\nu\in\mathbb{F}_2^n}\big|W_F(\mu,\nu)\big|.
\end{align*}

Let $f$ be a Boolean function on $\mathbb{F}_2^n$. Then it is easy to check that
\begin{align}\label{eqparseval}
\sum_{\nu\in\mathbb{F}_2^n}W_f^2(\nu)=2^{2n},
\end{align}
which is the well-known {\it Parseval formula}. It indicates that
$\max_{\nu\in\mathbb{F}_2^n}\big|W_f(\nu)\big|\geq 2^{n/2}$. The equality occurs if and only if $n$ is even and $W_f(\nu)=\pm 2^{n/2}$ for any $\nu\in\mathbb{F}_2^n$. In this case, $f$ is called a {\it bent function}. It is well known that every bent function $f$ satisfies that $W_f(\nu)=2^{n/2}(-1)^{f^*(\nu)}$ for any $\nu\in\mathbb{F}_2^n$, where $f^*$ is called the {\it dual} of $f$.

 According to Parseval formula, we can obtain the following result.

\begin{lemma}\label{lemwalsh<0}
Let $f$ be a Boolean function on $\mathbb{F}_2^n$. Then $W_f(\nu)\leq 0~({\rm or}~\geq 0)$ for any $\nu\in \mathbb{F}_2^n$ if and only if $f$ is affine with $f({\bf0_n})=1~({\rm or}~0)$, where ${\bf 0_n}$ is the zero vector of $\mathbb{F}_2^n$.
\end{lemma}
\begin{proof}
Since the proof of the case ``$\geq 0$" is similar as that of ``$\leq 0$", we only prove the latter case here.

Let $f$ be affine with $f({\bf0_n})=1$, that is, $f(x)=\nu_0\cdot x+1$ for some vector $\nu_0\in\mathbb{F}_2^{n}$. Then it is easily obtained by definition that $W_f(\nu)$ is equal to $-2^n$ if $\nu=\nu_0$, and equal to 0 otherwise.

Conversely, suppose that $W_f(\nu)\leq 0$ for any $\nu\in\mathbb{F}_2^n$.
 Then by the relation
\begin{align}\label{eqinverse0}
\sum_{\nu\in\mathbb{F}_{2}^{n}}W_f(\nu)=\sum_{x\in\mathbb{F}_2^n}(-1)^{f(x)}\sum_{\nu\in\mathbb{F}_{2^n}}(-1)^{\nu\cdot x}=2^{n}(-1)^{f({\bf0_n})},
\end{align}
one has $\sum_{\nu\in\mathbb{F}_2^n}W_f(\nu)=-2^n$ and $f({\bf0_n})=1$. Combining Parseval formula, it holds
\begin{align*}
\bigg(\sum_{\nu\in\mathbb{F}_2^n}W_f(\nu)\bigg)^2=\sum_{\nu\in\mathbb{F}_2^n}W_f^2(\nu),
\end{align*}
which implies that $W_f(\mu)W_f(\nu)=0$ for any distinct vectors $\mu,\nu\in\mathbb{F}_2^n$. Equivalently, there is a unique $\nu_0\in\mathbb{F}_2^n$ such that $W_f(\nu_0)\neq 0$. Hence $W_f(\nu)$ equals $-2^n$ if $\nu=\nu_0$, and equals 0 otherwise. Therefore, $f(x)=\nu_0\cdot x+1$.
\end{proof}

It is well known \cite{DBLP:books/cu/10/CarletCH10a} that every $(n,m)$-function $F$ can be uniquely represented as a polynomial of the form
\begin{align}\label{ANF}
F(x)=\sum_{I\subseteq\{1,2,\ldots,n\}}a_I \bigg(\prod_{i\in I}x_i\bigg), \hspace{0.2cm}a_I\in\mathbb{F}_2^m,
\end{align}
which is usually called the {\it algebraic normal form (ANF)} of $F$. The {\it algebraic degree} of $F$, denoted by $\deg(F)$, is defined as the maximum cardinality of $I$ such that $a_I\neq{\bf 0_m}$ in its ANF. An $(n,m)$-function $F$ is called {\it affine} if $\deg(F)\leq 1$.

\begin{definition}\label{defplateaued}
A Boolean function $f$ on $\mathbb{F}_2^n$ is called {\it plateaued} if its Walsh-Hadamard transform satisfies that $W_f(\nu)\in\{0,\pm \Lambda\}$ for any $\nu\in\mathbb{F}_2^n$, where $\Lambda=2^{(n+\lambda)/2}$ for some $0\leq\lambda\leq n$ with the same parity of $n$ is called the {\it amplitude} of $f$. In particular, when $n$ is even and $\Lambda=2^{n/2}$, $f$ is clearly a bent function.
\end{definition}

\begin{definition}
An $(n,m)$-function $F$ is called {\it vectorial plateaued} if all its components are plateaued functions. In particular, $F$ is called {\it vectorial bent} if $n$ is even and $F$ is a vectorial plateaued function with single amplitude $2^{n/2}$; and $F$ is called {\it almost bent} (AB) if $n$ is odd and $F$ is a vectorial plateaued function with single amplitude $2^{(n+1)/2}$.
\end{definition}

\subsection{Minimal linear codes}

For a vector $c=(c_1,c_2,\ldots,c_n)\in\mathbb{F}_2^n$, the set $\{1\leq i\leq n : c_i=1\}$ (denoted by ${\rm suppt}(c)$) is called the {\it support} of $c$, whose cardinality (usually denoted by ${\rm wt}(c)$) is called the {\it (Hamming) weight} of $c$, i.e., ${\rm wt}(c)=|{\rm suppt}(c)|=\sum_{i=1}^n c_i$. Any vector $c^{(1)}\in\mathbb{F}_2^n$ is said to be {\it covered} by $c^{(2)}\in\mathbb{F}_2^n$, denoted  by $c^{(1)}\preceq c^{(2)}$, if ${\rm suppt}(c^{(1)})\subseteq{\rm suppt}(c^{(2)})$. The {\it(Hamming) distance} of $c^{(1)},c^{(2)}\in\mathbb{F}_2^n$, denoted by $d(c^{(1)},c^{(2)})$, is defined as the weight of $c^{(1)}+c^{(2)}$.

Any code $C\subseteq \mathbb{F}_2^n$ is said to be an $[n,k,d]$ linear code if it is a $k$-dimensional linear subspace of $\mathbb{F}_2^n$ with minimal distance $d=\min \{d(c^{(1)},c^{(2)}) : c^{(1)},c^{(2)}\in C^*, c^{(1)}\neq c^{(2)}\}$, where $n$, $k$ and $d$ are called the {\it length, dimension and minimal distance} of $C$, respectively. Since $C$ is a linear subspace, $d$ is in fact the minimal weight of  all nonzero codewords in $C$.  The {\it weight enumerator} of $C$ is defined by a polynomial
$1+A_1z+A_2z^2+\cdots+A_nz^n$, where each coefficient $A_i$ denotes the number of codewords in $C$ whose weight is equal to $i$. The code $C$ is said to be a {\it $t$-weight} code if the number of nonzero coefficients $A_i$ in its weight enumerator is equal to $t$.

For a code $C$ over $\mathbb{F}_2$, a codeword $c\in C$ is called {\it minimal} if in $C$ it is covered by itself only. A linear code $C$ is called minimal if every nonzero codeword in $C$ is minimal. The following result is a sufficient condition for a linear code to be minimal.

\begin{lemma}\cite{AB-condition-1998}\label{lemABcondition}
Let $C\subseteq \mathbb{F}_2^n$ be a linear code. Let $w_{\max}$ and $w_{\min}$ be the maximum and minimum  nonzero  Hamming weights in $C$, respectively. If
\begin{align}\label{eqAB-condition}
\frac{w_{\min}}{w_{\max}}> \frac{1}{2},
\end{align}
then $C$ is minimal.
\end{lemma}

\begin{remark}\label{characp}
If $C$ is a linear code over $\mathbb{F}_p$, where $p$ is a prime, then \eqref{eqAB-condition} in Lemma \ref{lemABcondition} becomes  that $\frac{w_{\min}}{w_{\max}}> \frac{p-1}{p}$.
\end{remark}

The condition \eqref{eqAB-condition} is called the {\it AB condition} in this paper. AB condition is a sufficient but not necessary condition for a linear code to be minimal. Until now, many infinite families of minimal linear codes  violating the AB condition have been found, see for instance \cite{Bonini-2020-JAC, Chang-2018-DCC,Ding-2018-IT,Lixia-qinyue-2020,Sihem-2020-IT,Pasalic-2021-CCDS,Zhangweiqin-2019-AAECC,Zhangfengrong-2021-DCC,Zhangfengrong-2022-DCC}. The following result gives a necessary and sufficient condition for a binary linear code to be minimal.

\begin{theorem}\cite{Ding-2018-IT}\label{thdingIT}
Let $C\subseteq \mathbb{F}_2^n$ be a binary linear code. Then $C$ is minimal if and only if
 \begin{align}\label{eqthding}
 {\rm wt}(c^{(1)}+c^{(2)})\neq{\rm wt}(c^{(2)})-{\rm wt}(c^{(1)}),
 \end{align}
 for any $c^{(1)}, c^{(2)}, c^{(1)}+c^{(2)}\in C^*$.
\end{theorem}

\section{A generic construction of minimal binary linear codes}\label{sec:generic1}

Recently, many minimal binary linear codes have been obtained from the linear code of the form (\ref{eqcodeBooleanintro}):
\begin{align*}
C_f=\bigg\{c(\mu,\nu)=\big(\mu f(x)+\nu \cdot x\big)_{x\in\mathbb{F}_{2}^{n*}} : \mu\in\mathbb{F}_2,\nu\in\mathbb{F}_{2}^{n}\bigg\}
\end{align*}
by choosing some suitable Boolean functions $f$ on $\mathbb{F}_{2^n}$ (see e.g., \cite{Bonini-2020-JAC,Ding-2016,Ding-2018-IT,Heng-2018-FFA,Sihem-2020-IT,Sihem-2020-IT2,Xu-2020-FFA}). In this section, we consider this linear code for the case $f$ being a vectorial Boolean function.

\begin{construction}\label{ourcon}
Let $F$ be an $(n,m)$-function. Define a linear code as in (\ref{eqourconintro}):
\begin{align*}
C_F=\bigg\{c(\mu,\nu)=\big(\mu\cdot F(x)+\nu\cdot x\big)_{x\in\mathbb{F}_{2}^{n*}} : \mu\in\mathbb{F}_2^m,\nu\in\mathbb{F}_2^n\bigg\}.
\end{align*}
\end{construction}

In fact, a more general linear code defined as
\begin{align}\label{eqcarletcon}
C_F=\bigg\{c(\mu,\nu,\lambda)=\big(\mu\cdot F(x)+\nu\cdot x+\lambda\big)_{x\in\mathbb{F}_{2}^{n}} : \mu\in\mathbb{F}_2^m,\nu\in\mathbb{F}_2^n,\lambda\in\mathbb{F}_2\bigg\}
\end{align}
 has been considered in \cite{Carlet-ding-2007-FFA} and \cite{Wadayama-2001}, where it was proved to be a $[2^n,n+m+1,NL(F)]$ binary linear code when $NL(F)\neq 0$ (see \cite[Proposition 6]{Carlet-ding-2007-FFA} and \cite[Lemma 10]{Wadayama-2001}), where $NL(F)$ is the nonlinearity of $F$. But this code is clearly not minimal, since $c({\bf0_m},{\bf0_n},1)\in C_F$ with ${\rm wt}(c({\bf0_m},{\bf0_n},1))=2^n$. Hence, in this section, we give a further study on the binary linear code $C_F$ in Construction \ref{ourcon} to find some minimal codes. We first give an observation on the parameters of $C_F$ in the following subsection.

\subsection{The parameters of $C_F$}

It is noticed that, the parameters (including the length, dimension and minimal distance) of the linear code $C_F$ in \eqref{eqcarletcon}  have been obtained in \cite{Carlet-ding-2007-FFA} and \cite{Wadayama-2001}, respectively. The code $C_F$ in Construction \ref{ourcon} (i.e., the code in \eqref{eqourconintro}) is obtained by modifying \eqref{eqcarletcon} slightly. So the parameters of the former can be obtained similarly as that of the later. For completeness, we also present the parameters of $C_F$ in Construction \ref{ourcon} in this subsection.
 The techniques used here are similar to that of  \cite{Bonini-2020-JAC,Ding-2018-IT,Wadayama-2001}.

\begin{proposition}\label{propdim}
Let $F$ be an $(n,m)$-function such that $\mu\cdot F(x)$ is not linear for any $\mu\in \mathbb{F}_2^{m*}$. Then the binary linear code $C_F$ generated by Construction \ref{ourcon} has length $2^n-1$ and dimension $n+m$.
\end{proposition}
\begin{proof}
Clearly, the length of $C_F$ is $2^n-1$.

It is noted that every codeword $c(\mu,\nu)=\big(\mu\cdot F(x)+\nu\cdot x\big)_{x\in\mathbb{F}_2^{n^*}}$ in $C_F$ can be linearly represented by $c(\mu^{(i)},{\bf0_n})=\big(\mu^{(i)}\cdot F(x)\big)_{x\in\mathbb{F}_2^{n^*}}$, $i=1,2,\ldots,m$, and $c({\bf0_m},\nu^{(j)})=(\nu^{(j)}\cdot x)_{x\in\mathbb{F}_2^{n^*}}$, $j=1,2,\ldots,n$, where $\{\mu^{(1)},\mu^{(2)},\ldots,\mu^{(m)}\}$ is a basis of $\mathbb{F}_2^m$ over $\mathbb{F}_2$, and $\{\nu^{(1)},\nu^{(2)},\ldots,\nu^{(n)}\}$ is a basis of $\mathbb{F}_2^n$ over $\mathbb{F}_2$.  Below, we prove that
these codewords are linearly independent over $\mathbb{F}_2$.

Assume that there are some $a_i, b_j\in \mathbb{F}_2$, $i=1, 2, \ldots, m$, $j=1, 2, \ldots, n$, such that
\begin{align*}
\sum_{i=1}^m a_i c(\mu^{(i)},{\bf0_n})+\sum_{j=1}^{n}b_j c({\bf0_m},\nu^{(j)})={\bf0}.
\end{align*}
Then we have
\begin{align}\label{eqpropdim1}
\mu'\cdot F(x)+\nu'\cdot x=0, \hspace{0.3cm}\forall~x\in\mathbb{F}_2^{n*},
\end{align}
where $\mu'=a_1\mu^{(1)}+a_2\mu^{(2)}+\cdots+a_m\mu^{(m)}$ and $\nu'=b_1\nu^{(1)}+b_2\nu^{(2)}+\cdots+b_n\nu^{(n)}$. Firstly, we claim that $\mu'={\bf0_m}$. Otherwise, \eqref{eqpropdim1} is impossible since $\mu'\cdot F(x)$ is not linear by the assumption. This also implies that $\nu'={\bf0_n}$, since the cardinality of $\{x\in\mathbb{F}_2^{n*} : \nu'\cdot x=1\}$ is $2^{n-1}$ when $\nu'\neq {\bf0_n}$.
 Hence, we obtain that $\mu'={\bf0_m}$ and $\nu'={\bf0_n}$, which implies that $a_i=0$ for any $1\leq i\leq m$, and $b_j=0$ for any $1\leq j\leq n$. This completes the proof.
\end{proof}

We now characterize the weight distribution of the binary linear code $C_F$ given in Construction \ref{ourcon}.

\begin{proposition}\label{propweightdistri}
Let $F$ be an $(n,m)$-function with $F({\bf0_n})={\bf0_m}$, and let $C_F$ be the binary linear code generated by Construction \ref{ourcon}. Then for any $c(\mu,\nu)\in C_F$, it holds that
\begin{align*}
{\rm wt}(c(\mu,\nu))=
\begin{cases}
0, \hspace{0.3cm}&{\rm if}~\mu={\bf0_m},\nu={\bf0_n};\\
2^{n-1}, &{\rm if}~\mu={\bf0_m},\nu\in\mathbb{F}_2^{n*};\\
2^{n-1}- \frac{1}{2}W_F(\mu,\nu), &{\rm otherwise.}
\end{cases}
\end{align*}
\end{proposition}
\begin{proof}
By definition, for any $(\mu,\nu)\in\mathbb{F}_2^m\times\mathbb{F}_2^n$, the weight of $c(\mu,\nu)\in C_F$ is given by
\begin{align}\label{eqpropweight}
{\rm wt}(c(\mu,\nu))=\sum_{x\in\mathbb{F}_2^{n*}}(\mu\cdot F(x)+\nu\cdot x)=\sum_{x\in\mathbb{F}_2^n}(\mu\cdot F(x)+\nu\cdot x).
\end{align}
Since $\mu\cdot F(x)+\nu\cdot x$ is a Boolean function, we have $\mu\cdot F(x)+\nu\cdot x=\frac{1-(-1)^{\mu\cdot F(x)+\nu\cdot x}}{2}$. Hence \eqref{eqpropweight} becomes that
\begin{align*}
{\rm wt}(c(\mu,\nu))=2^{n-1}-\frac{1}{2}\sum_{x\in\mathbb{F}_2^n}(-1)^{\mu\cdot F(x)+\nu\cdot x},
\end{align*}
which is equal to 0 if $\mu={\bf0_m}, \nu={\bf0_n}$; equal to $2^{n-1}$ if $\mu={\bf0_m},\nu\neq{\bf0_n}$; and equal to $2^{n-1}-\frac{1}{2}W_{F}(\mu,\nu)$ otherwise.
\end{proof}

\begin{remark}\label{rmkdistance}
When $F$ is not affine, Lemma \ref{lemwalsh<0} implies that $\max_{\mu\in\mathbb{F}_2^{m*},\nu\in\mathbb{F}_2^n}W_F(\mu,\nu)>0$. Thus, by
 Proposition \ref{propweightdistri}, we obtain that the minimal distance $d$ of the binary linear code $C_F$ in Construction \ref{ourcon} satisfies that
\begin{align*}
d=2^{n-1}-\frac{1}{2}\max_{\mu\in\mathbb{F}_2^{m*},\nu\in\mathbb{F}_2^n}\big|W_F(\mu,\nu)\big|=NL(F).
\end{align*}
\end{remark}

\begin{remark}\label{rmk3weight}
When $F$ is a vectorial plateaued function with single amplitude, the linear code $C_F$ in Construction \ref{ourcon} is a 3-weight linear code by Proposition \ref{propweightdistri}.
\end{remark}

\begin{remark}\label{rmkabcondition}
Let $F$ be an $(n,m)$-function such that $\mu\cdot F(x)$ is not affine for some $\mu\in\mathbb{F}_{2}^{m*}$. Then, similarly as the proof of \cite[Proposition 3.2]{Pasalic-2021-CCDS}, one can show that the linear code $C_F$ in Construction \ref{ourcon} does not satisfy the AB condition if and only if
\begin{align}\label{eqrmkabcondition}
2\max_{\mu\in\mathbb{F}_2^{m*},\nu\in\mathbb{F}_2^n}W_F(\mu,\nu)-\min_{\mu\in\mathbb{F}_2^{m*},\nu\in\mathbb{F}_2^n}W_F(\mu,\nu)\geq 2^n.
\end{align}
\end{remark}

Reformulating the above results to the finite field version, we have the following result.

\begin{proposition}\label{propfieldversion}
Let $F$ be a function from $\mathbb{F}_{2^n}$ to $\mathbb{F}_{2^m}$ such that $F({0})={0}$, and ${\rm Tr}_1^m(\mu F(x))$ is not affine for any $\mu\in\mathbb{F}_{2^m}^*$. Let $C_F$ be the binary linear code defined as
\begin{align}\label{eqourconfield}
C_F=\bigg\{c(\mu,\nu)=\big({\rm Tr}_1^m(\mu F(x))+{\rm Tr}_1^n(\nu x)\big)_{x\in\mathbb{F}_{2^n}^*} : (\mu,\nu)\in\mathbb{F}_{2^m}\times\mathbb{F}_{2^n}\bigg\}.
\end{align}
Then $C_F$ is a linear code with parameters $[2^n-1,n+m,NL(F)]$.
\end{proposition}

\subsection{Minimality of $C_F$}

In this subsection, we give some conditions for the binary linear code $C_F$ in Construction \ref{ourcon} to be minimal. The main result is given as follows.

\begin{theorem}\label{thmainresult}
Let $F$ be an $(n,m)$-function such that $F({\bf0_n})={\bf0_m}$ and $\mu\cdot F$ is not affine for any $\mu\in\mathbb{F}_2^{m*}$. Let $C_F$ be the linear code generated by Construction \ref{ourcon}. Then $C_F$ is minimal if and only if both the following two conditions are satisfied:
\begin{enumerate}
\item[(1)] For any $\mu\in\mathbb{F}_2^{m*}$ and any $\nu,\nu'\in\mathbb{F}_2^n$ with $\nu\neq \nu'$, it holds that
\begin{align}\label{eqthmainresult1}
W_F(\mu,\nu)\pm W_F(\mu,\nu')\neq 2^n;
\end{align}

\item[(2)] For any $\mu,\mu'\in\mathbb{F}_{2}^{m*}$ with $\mu\neq\mu'$, and any $\nu,\nu'\in\mathbb{F}_2^n$, it holds that
\begin{align}\label{eqthmainresult2}
W_F(\mu,\nu)+W_F(\mu',\nu')-W_F(\mu+\mu',\nu+\nu')\neq 2^n.
\end{align}
\end{enumerate}
\end{theorem}
\begin{proof}
 By the definition of $C_F$ and the assumption that $\mu\cdot F$ is not affine for any $\mu\in\mathbb{F}_2^{m*}$,  it is easily seen that $c(\mu,\nu)={\bf0}$ if and only if $(\mu,\nu)=({\bf0_m},{\bf0_n})$, and $c(\mu,\nu)+c(\mu',\nu')=c(\mu+\mu',\nu+\nu')$ for any $(\mu,\nu), (\mu',\nu')\in\mathbb{F}_2^m\times \mathbb{F}_2^n$. So by theorem \ref{thdingIT}, we  need to show that
\begin{align}\label{eqproofthmainr1}
{\rm wt}\big(c(\mu+\mu',\nu+\nu')\big)\neq {\rm wt}\big(c(\mu,\nu)\big)-{\rm wt}\big(c(\mu',\nu')\big),
\end{align}
for any $(\mu, \nu), (\mu', \nu'), (\mu+\mu', \nu+\nu')\in\mathbb{F}_2^m\times\mathbb{F}_2^n\backslash\{({\bf0_m}, {\bf0_n})\}$. According to whether $\mu$ and $\mu'$ are nonzero, our proof can be divided into the following five cases.

{\bf Case 1}. $\mu=\mu'={\bf 0_m}$ and $\nu, \nu', \nu+\nu'\in\mathbb{F}_{2}^{n*}$. In this case, \eqref{eqproofthmainr1} obviously holds as ${\rm wt}(c(\mu,\nu))={\rm wt}(c(\mu',\nu'))={\rm wt}(c(\mu+\mu',\nu+\nu'))=2^{n-1}$ by Proposition \ref{propweightdistri}.

{\bf Case 2}. $\mu={\bf0_m}, \mu'\neq {\bf 0_m}$ and $\nu\neq{\bf0_n}$. In this case, ${\rm wt}(c(\mu,\nu))=2^{n-1}$, ${\rm wt}(c(\mu',\nu'))=2^{n-1}-\frac{1}{2}W_F(\mu',\nu')$, and ${\rm wt}(c(\mu+\mu',\nu+\nu'))=2^{n-1}-\frac{1}{2}W_F(\mu',\nu+\nu')$ by Proposition \ref{propweightdistri}. Hence, \eqref{eqproofthmainr1} holds if and only if
\begin{align*}
W_F(\mu',\nu')+W_F(\mu',\nu+\nu')\neq 2^n.
\end{align*}

{\bf Case 3}. $\mu\neq {\bf 0_m}, \mu'={\bf 0_m}$ and $\nu'\neq {\bf 0_n}$. In this case, ${\rm wt}(c(\mu,\nu))=2^{n-1}-\frac{1}{2}W_F(\mu,\nu)$, ${\rm wt}(c(\mu',\nu'))=2^{n-1}$, ${\rm wt}(c(\mu+\mu',\nu+\nu'))=2^{n-1}-\frac{1}{2}W_F(\mu,\nu+\nu')$ by Proposition \ref{propweightdistri}. Hence, \eqref{eqproofthmainr1} holds if and only if
\begin{align*}
W_F(\mu,\nu+\nu')-W_F(\mu,\nu)\neq 2^n.
\end{align*}

{\bf Case 4}. $\mu=\mu'\neq {\bf 0_m}$ and $\nu\neq\nu'$. In this case,  ${\rm wt}(c(\mu, \nu))=2^{n-1}-\frac{1}{2}W_F(\mu, \nu)$, ${\rm wt}(c(\mu', \nu'))=2^{n-1}-\frac{1}{2}W_F(\mu', \nu')$, ${\rm wt}(c(\mu+\mu', \nu+\nu'))=2^{n-1}$ by Proposition \ref{propweightdistri}. Hence, \eqref{eqproofthmainr1} holds if and only if
\begin{align*}
W_F(\mu,\nu')-W_F(\mu,\nu)\neq 2^n.
\end{align*}

{\bf Case 5}. $\mu\neq {\bf 0_m}, \mu'\neq {\bf 0_m}$ and $\mu+\mu'\neq{\bf 0_m}$. In this case, ${\rm wt}(c(\mu, \nu))=2^{n-1}-\frac{1}{2}W_F(\mu, \nu)$, ${\rm wt}(c(\mu', \nu'))=2^{n-1}-\frac{1}{2}W_F(\mu', \nu')$, ${\rm wt}(c(\mu+\mu', \nu+\nu'))=2^{n-1}-\frac{1}{2}W_F(\mu+\mu', \nu+\nu')$ by Proposition \ref{propweightdistri}. Hence, \eqref{eqproofthmainr1} holds if and only if
\begin{align*}
W_F(\mu', \nu')-W_F(\mu, \nu)+W_F(\mu+\mu', \nu+\nu')\neq 2^n.
\end{align*}
The result follows then from the above five cases.
\end{proof}

\begin{remark}
When $m=1$, that is, $F$ is a Boolean function, Theorem \ref{thmainresult} reduces to  \cite[Theorem 15]{Ding-2018-IT}.
\end{remark}

\section{Three-weight minimal linear codes from vectorial plateaued functions}\label{sec:threeweight}

In this section, we present some three-weight minimal linear codes from certain vectorial plateaued functions. The main result is given as follows.

\begin{theorem}\label{thfplateaued}
Let $n>4$. Let $F$ be an  $(n,m)$-function satisfying the following two conditions:
\begin{enumerate}
\item[(1)] $F({\bf 0_n})={\bf 0_m}$;

\item[(2)] $\max_{\mu\in\mathbb{F}_2^{m*},\nu\in\mathbb{F}_2^n} |W_F(\mu,\nu)|=2^{\frac{n+\lambda}{2}}$ such that $0\leq\lambda\leq n-4$.
\end{enumerate}
Then the code $C_F$ generated by Construction \ref{ourcon} is a minimal binary linear code  with parameters $[2^n-1,n+m,2^{n-1}-2^{\frac{n+\lambda}{2}-1}]$.
\end{theorem}
\begin{proof}
Firstly, $C_F$ is obviously a binary linear code with  parameters $[2^n-1,n+m,2^{n-1}-2^{\frac{n+\lambda}{2}-1}]$  by Proposition \ref{propdim} and Remark \ref{rmkdistance}, since $NL(F)=2^{n-1}-2^{\frac{n+\lambda}{2}-1}\neq 0$ (which means that $F$ has no affine components).

 Then, we only need to show that $C_F$ is minimal, which can be done by verifying the two conditions of Theorem \ref{thmainresult}. In fact, both conditions are obvious, since for any $\mu\in\mathbb{F}_2^{m*}$ and any distinct $\nu,\nu'\in\mathbb{F}_{2}^n$, one has
\begin{align*}
|W_F(\mu,\nu)\pm W_F(\mu,\nu')|\leq 2^{\frac{n+\lambda}{2}+1},
\end{align*}
which is strictly less than $2^n$ when $\lambda\leq n-4$; and for any distinct $\mu, \mu'\in\mathbb{F}_2^{m*}$, and any $\nu,\nu'\in\mathbb{F}_{2}^n$, one has
\begin{align*}
|W_F(\mu,\nu)+W_F(\mu',\nu')-W_F(\mu+\mu',\nu+\mu')|\leq  3\times 2^{\frac{n+\lambda}{2}},
\end{align*}
which is also strictly less than $2^n$ when $\lambda\leq n-4$.
\end{proof}

Applying Theorem \ref{thfplateaued} to vectorial plateaued functions with single amplitude, we obtain the following result.

\begin{corollary}\label{corsingleplateaued}
Let $n>4$ and $0\leq \lambda\leq n-4$. Let $F$ be a vectorial plateaued $(n,m)$-function such that $F({\bf 0_n})={\bf 0_m}$ and $F$ has the single amplitude $2^{\frac{n+\lambda}{2}}$.
Then the code $C_F$ defined in Construction \ref{ourcon} is a three-weight minimal linear code with parameters $[2^n-1,n+m,2^{n-1}-2^{\frac{n+\lambda}{2}-1}]$. Moreover, the weight distribution of $C_F$ is given in Table \ref{b2}.
\end{corollary}
\begin{table*}[!htbp]\setlength{\abovecaptionskip}{0cm}
\caption{Weight distribution of $C_F$ in Corollary \ref{corsingleplateaued}}  \centering \label{b2}
\medskip
\begin{tabular}{|c|c|}
 \hline
 Weight & Frequency\\
 \hline
 0 & 1\\
 \hline
 $2^{n-1}$ & $2^{n}-1+(2^m-1)(2^n-2^{n-\lambda})$\\
 \hline
 $2^{n-1}+2^{\frac{n+\lambda}{2}-1}$ & $(2^m-1)(2^{n-\lambda-1}-2^{\frac{n-\lambda}{2}-1})$\\
 \hline
 $2^{n-1}-2^{\frac{n+\lambda}{2}-1}$ & $(2^m-1)(2^{n-\lambda-1}+2^{\frac{n-\lambda}{2}-1})$\\
 \hline
\end{tabular}
\end{table*}

\begin{proof}
The first part of this corollary is clear by Theorem \ref{thfplateaued} and Remark \ref{rmk3weight}. Below we calculate the weight distribution of $C_F$.

Since $W_F(\mu,\nu)\in\{0,\pm 2^{\frac{n+\lambda}{2}}\}$ for any $(\mu,\nu)\in\mathbb{F}_2^{m*}\times\mathbb{F}_2^n$,  we have that
$${\rm wt}(\mu,\nu)\in T:=\{0, 2^{n-1}, 2^{n-1}+2^{\frac{n+\lambda}{2}-1},2^{n-1}-2^{\frac{n+\lambda}{2}-1}\}$$
 for any $(\mu,\nu)\in\mathbb{F}_2^{m}\times\mathbb{F}_2^n$ by proposition \ref{propweightdistri}. To determine the weight distribution of $C_F$, we need to determine the frequency of every element in $T$, which are denoted by $\mathbf{f}(0)$, $\mathbf{f}(2^{n-1})$, $\mathbf{f}(2^{n-1}+2^{\frac{n+\lambda}{2}-1})$ and $\mathbf{f}(2^{n-1}-2^{\frac{n+\lambda}{2}-1})$, respectively.

For every fixed $\mu\in\mathbb{F}_2^{m*}$, let
$
S_F:=\{\nu\in\mathbb{F}_2^n : W_F(\mu,\nu)\neq 0\}.
$
Then by Parseval formula $\sum_{\nu\in\mathbb{F}_2^n}W_F^2(\mu,\nu)=2^{2n}$, it is easily obtained that $|S_F|=2^{n-\lambda}$. Hence $|\overline{S_F}|=2^n-2^{n-\lambda}$, where $\overline{S_F}=\mathbb{F}_2^n\backslash S_F$. Let
\begin{align*}
S_F^{(+)}:=\{\nu\in\mathbb{F}_2^n : W_F(\mu,\nu)>0\} \hspace{0.3cm}{\rm and}\hspace{0.3cm} S_F^{(-)}:=\{\nu\in\mathbb{F}_2^n : W_F(\mu,\nu)<0\},
\end{align*}
then we have
\begin{align}\label{eqcorweightp1}
|S_F^{(+)}|+|S_F^{(-)}|=2^{n-\lambda}.
\end{align}
In addition, by \eqref{eqinverse0}, we get
\begin{align*}
\sum_{\nu\in\mathbb{F}_2^n}W_F(\mu,\nu)=2^n(-1)^{\mu\cdot F({\bf 0_n})}=2^n.
\end{align*}
Consequently, we have
\begin{align}\label{eqcorweightp2}
2^{\frac{n+\lambda}{2}}\big(|S_F^{(+)}|-|S_F^{(-)}|\big)=2^n.
\end{align}
Combining \eqref{eqcorweightp1} and \eqref{eqcorweightp2}, we obtain that $|S_F^{(+)}|=\frac{1}{2}(2^{n-\lambda}+2^{\frac{n-\lambda}{2}})$  and $|S_F^{(-)}|=\frac{1}{2}(2^{n-\lambda}-2^{\frac{n-\lambda}{2}})$.

Therefore, by proposition \ref{propweightdistri}, we have
\begin{align*}
&\mathbf{f}(0)=1, \hspace{0.3cm} \mathbf{f}(2^{n-1})=2^{n}-1+(2^m-1)|\overline{S_F}|=2^{n}-1+(2^m-1)(2^n-2^{n-\lambda}),\\
&\mathbf{f}(2^{n-1}+2^{\frac{n+\lambda}{2}-1})=(2^m-1)|S_F^{(-)}|=(2^m-1)(2^{n-\lambda-1}-2^{\frac{n-\lambda}{2}-1}), \hspace{0.3cm} {\rm and}\\
& \mathbf{f}(2^{n-1}-2^{\frac{n+\lambda}{2}-1})=(2^m-1)|S_F^{(+)}|=(2^m-1)(2^{n-\lambda-1}+2^{\frac{n-\lambda}{2}-1}).
\end{align*}
This completes the proof.
\end{proof}

Applying Corollary \ref{corsingleplateaued} to vectorial bent functions and AB functions, respectively, we obtain the following two results.

\begin{corollary}\label{corVB}
Let $n>4$ be an even integer and let $m\leq \frac{n}{2}$. Let $F$ be a vectorial bent $(n,m)$-function with $F({\mathbf{0_n}})=\mathbf{0_m}$. Then the code $C_F$ generated by Construction \ref{ourcon} is a three-weight minimal linear code with parameters $[2^n-1, n+m, 2^{n-1}-2^{\frac{n}{2}-1}]$. Moreover, the weight distribution of $C_F$ is given in Table \ref{b3}.
\end{corollary}

\begin{table*}[!htbp]\setlength{\abovecaptionskip}{0cm}
\caption{Weight distribution of $C_F$ in Corollary \ref{corVB}}  \centering \label{b3}
\medskip
\begin{tabular}{|c|c|}
 \hline
 Weight & Frequency\\
 \hline
 0 & 1\\
 \hline
 $2^{n-1}$ & $2^{n}-1$\\
 \hline
 $2^{n-1}+2^{\frac{n}{2}-1}$ & $(2^m-1)(2^{n-1}-2^{\frac{n}{2}-1})$\\
 \hline
 $2^{n-1}-2^{\frac{n}{2}-1}$ & $(2^m-1)(2^{n-1}+2^{\frac{n}{2}-1})$\\
 \hline
\end{tabular}
\end{table*}

\begin{remark}
For any even integer $n>4$, the minimal distance of the linear codes $C_F$ in Corollary \ref{corVB} reaches the optimal amongst the linear codes in Construction \ref{ourcon}, since their nonlinearity is the highest.
\end{remark}

\begin{corollary}\label{corAB}
Let $n>4$ be an odd positive integer. Let $F$ be an AB function with $F({\mathbf{0_n}})=\mathbf{0_n}$. Then the code $C_F$ generated by Construction \ref{ourcon} is a three-weight minimal linear code with parameters $[2^n-1,2n,2^{n-1}-2^{\frac{n+1}{2}-1}]$. Moreover, the weight distribution of $C_F$ is given in Table \ref{b4}.
\end{corollary}

\begin{table*}[!htbp]\setlength{\abovecaptionskip}{0cm}
\caption{Weight distribution of $C_F$ in Corollary \ref{corAB}}  \centering \label{b4}
\medskip
\begin{tabular}{|c|c|}
 \hline
 Weight & Frequency\\
 \hline
 0 & 1\\
 \hline
 $2^{n-1}$ & $2^{n}-1+(2^n-1)2^{n-1}$\\
 \hline
 $2^{n-1}+2^{\frac{n+1}{2}-1}$ & $(2^n-1)(2^{n-2}-2^{\frac{n-1}{2}-1})$\\
 \hline
 $2^{n-1}-2^{\frac{n+1}{2}-1}$ & $(2^n-1)(2^{n-2}+2^{\frac{n-1}{2}-1})$\\
 \hline
\end{tabular}
\end{table*}

The following numerical data is consistent with the conclusions of Corollaries \ref{corVB} and \ref{corAB}, respectively.

\begin{example}\label{exa1}
Let $n=6$ and $m=3$. Then the code $C_F$ in Corollary \ref{corVB} is a three-weight minimal linear code with parameters $[63,9,28]$ and weight enumerator
\begin{align*}
1+252z^{28}+63 z^{32}+196z^{36}.
\end{align*}
\end{example}

\begin{example}\label{exa2}
Let $n=7$. Then the code $C_F$ in Corollary \ref{corAB} is a three-weight minimal linear code with parameters $[127,14,56]$ and weight enumerator
\begin{align*}
1+4572z^{56}+8255 z^{64}+3556z^{72}.
\end{align*}
\end{example}

Note that \cite[Examples 20, 24, 33]{Ding-2018-IT} are also examples of minimal linear codes for $n=6$ and $n=7$. According to the values of $n$, we list the parameters of linear codes in Examples \ref{exa1}, \ref{exa2}, and \cite[Examples 20, 24, 33]{Ding-2018-IT} in Tables \ref{bexa1} and \ref{bexa2}.

\begin{table*}[!htbp]\setlength{\abovecaptionskip}{0cm}
\caption{The parameters of linear codes in Example \ref{exa1}  and \cite[Example 20]{Ding-2018-IT}}  \centering \label{bexa1}
\medskip
\begin{tabular}{|c|c|c|c|c|}
 \hline
n=6& Length& Dimension & Minimal distance & Weights\\
 \hline
 Example \ref{exa1} & 63& 9 & 28 & 3-weight \\
 \hline
 \cite[Example 20]{Ding-2018-IT} & 63 &7 & 14 & 4-weight\\
 \hline
\end{tabular}
\end{table*}

\begin{table*}[!htbp]\setlength{\abovecaptionskip}{0cm}
\caption{The parameters of linear codes in Example \ref{exa2}  and \cite[Examples 24, 33]{Ding-2018-IT}}  \centering \label{bexa2}
\medskip
\begin{tabular}{|c|c|c|c|c|}
 \hline
n=7&Length & Dimension & Minimal distance & Weights\\
 \hline
 Example \ref{exa2} & 127 & 14 & 56 & 3-weight \\
 \hline
 \cite[Example 24]{Ding-2018-IT} & 127 & 8 & 52 & 6-weight\\
 \hline
 \cite[Example 33]{Ding-2018-IT} & 127 &8 & 28 & 6-weight\\
 \hline
\end{tabular}
\end{table*}

From Tables \ref{bexa1} and \ref{bexa2}, it is clear that the  dimension, minimal distance and weights of the minimal linear codes in Examples \ref{exa1}, \ref{exa2} are better than that of the minimal linear codes in \cite[Examples 20, 24, 33]{Ding-2018-IT}.

It is noted that $w_{\min}=2^{n-1}-2^{\frac{n+\lambda}{2}-1}$ and  $w_{\max}=2^{n-1}+2^{\frac{n+\lambda}{2}-1}$ in Theorem \ref{thfplateaued}, which satisfy $\frac{w_{\min}}{w_{\max}}>\frac{1}{2}$ when $\lambda\leq n-4$. Hence, the minimal linear codes obtained in this section satisfy the AB condition.

Linear codes with few weights play a significant role in data storage systems, communication systems and consumer electronics. Many papers are devoted to constructing such linear codes. At the end of this section, we collect the parameters of some known three-weight linear codes over $\mathbb{F}_2$ in Table \ref{b3-weight}.

\begin{table*}[!htbp]\setlength{\abovecaptionskip}{0cm}
 \small
\caption{The parameters of some known three-weight linear codes over $\mathbb{F}_2$}  \centering \label{b3-weight}
\medskip
\begin{tabular}{|c|c|c|c|}
 \hline
Length& Dimension & Minimal distance & References\\
 \hline
 $2^n-1$ & $n+1$ & $2^{n-1}-2^{\frac{n}{2}-1}$~or~ $2^{n-1}-2^{\frac{n}{2}-1}-1$ & \cite[Remark 19]{Ding-2018-IT} \\
 \hline
 $2^{n-1}-1$ & $n$ & $2^{n-2}-2^{\frac{n-3}{2}}$ & \cite[Theorem 7]{Ding-2015-IT2} \\
 \hline
 $n_f$ (see (19) of \cite{Ding-2015-IT2}) & $m$ & $\frac{n_f-2^{(m-1)/2}}{2}$ & \cite[Corollary 11]{Ding-2015-IT2} \\
 \hline
 $n_f$ (see (22) of \cite{Ding-2015-IT2}) & $m$ & $\frac{n_f-2^{m-1-r_f/2}}{2}$ & \cite[Theorem 14]{Ding-2015-IT2} \\
 \hline
 $2^n-1$ &   $n+1$ & $2^{n-1}+\theta_2$, where $\theta_2<0$, see \cite[Lemma 6]{taoran-2021-IT} & \cite[Remark 17]{taoran-2021-IT}  \\
 \hline
 $2^n-1$ &   $n+1$ & $2^{n-1}-1$ or  $2^{n-1}-2$ & {\cite[Remarks 1, 2]{Chang-2018-DCC}}  \\
 \hline
$2^{n-1}$ &   $n$ &  $2^{n-2}-2^{\frac{n-2}{2}}$ or $2^{n-2}-2^{\frac{n-3}{2}}$ & { \cite[Theorems 9, 21]{Ding-2016} } \\
 \hline
  $2^{n}-3\times 2^{\frac{n-1}{2}}+1$ &   $n$ &  $2^{n-1}-3\times 2^{\frac{n-3}{2}}$  & {\cite[Theorem 7]{Tangdeng-2020} } \\
 \hline
  $2^{n}$ &   $n+1$ &  $2^{n-1}-2^{\frac{s}{2}-1}2^{\frac{r+1}{2}}$, where $n=r+s$  & { \cite{Zhangfengrong-2022-DCC} } \\
 \hline
  $2^{n}-1$ &   $n+1$ &  $2^{n-1}-2^{\frac{n}{2}-1}$   & {\cite{Sihem-2017} } \\
 \hline
 $2^{n}-1$ &   $n+1$ &  $2^{n-1}-2^{\frac{n+s}{2}-1}$   & { \cite[Theorem 1]{Sihem-2019} } \\
 \hline
  $2^{n-1}-2^{\frac{n}{2}-1}-1$ &   $n$ &  $2^{n-2}-(3l-1)2^{\frac{n}{2}-2}$, where  $l<\frac{2^{n/2}+1}{3}$  & { \cite[Theorem 1 (i)]{Luogj-2018-2} } \\
 \hline
  $2^{2l}-1, ~n=3l$ &   $n$ &  $2^{2l-1}-2^{l-1}$  & { \cite[Theorem 1]{Cheny-2020-FFA} } \\
 \hline
  $2^{n}-1$ &   $n+m$ &  $2^{n-1}-2^{\frac{n+\lambda}{2}-1}$  & Corollary \ref{corsingleplateaued} \\
 \hline
   $2^{n}-1$ &   $n+m$ &  $2^{n-1}-2^{\frac{n}{2}-1}$  & Corollary \ref{corVB} \\
 \hline
  $2^{n}-1$ &   $2n$ &  $2^{n-1}-2^{\frac{n+1}{2}-1}$  & Corollary \ref{corAB} \\
 \hline
\end{tabular}
\end{table*}

From Table \ref{b3-weight}, it is easily seen that the dimensions and minimal distances of the linear codes from Corollaries \ref{corsingleplateaued}, \ref{corVB} and \ref{corAB} are better than many of the others.

\section{A generic construction of minimal linear codes violating the AB condition}\label{sec:generic2}

In this section, we  present a generic construction of minimal linear codes violating the AB condition, which is given as follows.
\begin{construction} \label{ourcon2}
 Let $f$ be a Boolean function on $\mathbb{F}_{2}^n$ with $f(\mathbf{0_n})=0$, and the following two conditions are satisfied:
 \begin{enumerate}
\item[{\rm({\bf a})}] $W_f(\nu)\pm W_f(\nu')\neq 2^n$ for any $\nu,\nu'\in\mathbb{F}_2^n$ with $\nu\neq \nu'$;

\item[{\rm({\bf b})}] $2\max_{\nu\in\mathbb{F}_2^n} W_f(\nu)-\min_{\nu\in\mathbb{F}_2^n} W_f(\nu)\geq 2^n$.
 \end{enumerate}
 Let $G$ be an $(n, m-1)$-function with $G(\mathbf{0_n})=\mathbf{0_{m-1}}$, such that $\mu_1f+\tilde{\mu}\cdot G$ is not affine for any nonzero $\mu=(\mu_1,\tilde{\mu})\in\mathbb{F}_2\times\mathbb{F}_2^{m-1}$,  and the linear code $C_G$ defined by
 \begin{align}\label{conG}
 C_G=\bigg\{c(\tilde{\mu},\nu)=\big(\tilde{\mu}\cdot G(x)+\nu\cdot x\big)_{x\in\mathbb{F}_2^{n*}} : (\tilde{\mu},\nu)\in\mathbb{F}_2^{m-1}\times \mathbb{F}_2^n\bigg\}
 \end{align}
 is minimal.
Let $F$ be an $(n,m)$-function given by
\begin{align}\label{mainfunction}
F(x)=(f(x), G(x)),
 \end{align}
 and let $C_F$ be the linear code generated by Construction \ref{ourcon}.
 \end{construction}

 In the following, we study the conditions for the linear code $C_F$ in Construction \ref{ourcon2} to be minimal and to be violated the AB condition. To this end, we need first to calculate the Walsh-Hadamard transform of $F$.

\begin{proposition}\label{propwalshtransform}
Let $F$ be the $(n,m)$-function defined in Construction \ref{ourcon2}. Then for any nonzero $\mu=(\mu_1,\tilde{\mu})\in\mathbb{F}_2\times\mathbb{F}_2^{m-1}$ and any $\nu\in\mathbb{F}_2^n$, it holds  that
\begin{align*}
W_F(\mu,\nu)=\begin{cases}
W_G(\tilde{\mu},\nu),\hspace{0.3cm}&{\rm if} ~\mu_1=0,~\tilde{\mu}\neq{\mathbf{0_{m-1}}};\\
W_f(\nu),\hspace{0.3cm}&{\rm if} ~\mu_1=1,~\tilde{\mu}={\mathbf{0_{m-1}}};\\
W_{A_{\tilde{\mu}}}(\nu), \hspace{0.3cm}&{\rm if} ~\mu_1=1,~\tilde{\mu}\neq{\mathbf{0_{m-1}}},
\end{cases}
\end{align*}
where $A_{{\tilde{\mu}}}(x)=f(x)+\tilde{\mu}\cdot G(x)$.
\end{proposition}
\begin{proof}
By the definition of  Walsh-Hadamard transform, one has
\begin{align*}
W_F(\mu,\nu)=\sum_{x\in\mathbb{F}_2^n}(-1)^{\mu\cdot F(x)+\nu\cdot x}=\sum_{x\in\mathbb{F}_2^n}(-1)^{\mu_1f(x)+\tilde{\mu}\cdot G(x)+\nu\cdot x}, ~~\forall~(\mu,\nu)\in\mathbb{F}_2^{m*}\times\mathbb{F}_2^n,
\end{align*}
from which the result follows clearly.
\end{proof}

Since it is required in Construction \ref{ourcon2} that $$2\max_{\nu\in\mathbb{F}_2^n} W_f(\nu)-\min_{\nu\in\mathbb{F}_2^n} W_f(\nu)\geq 2^n,$$ by Proposition \ref{propwalshtransform} one immediately gets $$2\max_{\mu\in\mathbb{F}_2^{m*},\nu\in\mathbb{F}_2^n}W_F(\mu,\nu)
-\min_{\mu\in\mathbb{F}_2^{m*},\nu\in\mathbb{F}_2^n}W_F(\mu,\nu)\geq 2^n.$$
Then by Remark \ref{rmkabcondition}, we arrive at the following result.

\begin{proposition}\label{propvoiAB}
Let $C_F$ be the linear code generated by Construction \ref{ourcon2}. Then $C_F$ does not satisfy the AB condition.
\end{proposition}

 Thus, we only need to investigate the conditions under which the linear code $C_F$ in Construction \ref{ourcon2} is minimal. We give the following theorem.

\begin{theorem}\label{thgenericAB}
The linear code $C_F$ generated by Construction \ref{ourcon2} is minimal if and only if the following three conditions are fulfilled:
 \begin{enumerate}
\item[(1)] For any $\tilde{\mu}\in\mathbb{F}_2^{m-1}\backslash\{\mathbf{0_{m-1}}\}$ and any $\nu,\nu'\in\mathbb{F}_2^n$ with $\nu\neq \nu'$, it holds that
\begin{align}\label{eqgAB1}
W_{A_{\tilde{\mu}}}(\nu)\pm W_{A_{\tilde{\mu}}}(\nu')\neq 2^{n};
\end{align}

\item[(2)] For any $\tilde{\mu}\in\mathbb{F}_2^{m-1}\backslash\{\mathbf{0_{m-1}}\}$ and any $\nu,\nu'\in\mathbb{F}_2^n$, it holds that
\begin{align}\label{eqgAB2}
(-1)^{i}W_f(\nu)+(-1)^{j}W_G(\tilde{\mu},\nu')+(-1)^kW_{A_{\tilde{\mu}}}(\nu+\nu')\neq 2^n,
\end{align}
where $i,j,k\in \{0, 1\}$ such that exactly two of them are 0;

\item[(3)] For any  $\tilde{\mu},\tilde{\mu}'\in\mathbb{F}_2^{m-1}\backslash\{\mathbf{0_{m-1}}\}$ with $\tilde{\mu}\neq\tilde{\mu}'$,  and any $\nu,\nu'\in\mathbb{F}_2^n$, it holds that
    \begin{align}\label{eqgAB3}
W_{A_{\tilde{\mu}}}(\nu)+(-1)^{i}W_G(\tilde{\mu}',\nu')+(-1)^{j}W_{A_{\tilde{\mu}+\tilde{\mu}'}}(\nu+\nu')\neq 2^n,
\end{align}
where $i, j\in \{0, 1\}$ such that exactly one of them is 0.
\end{enumerate}
\end{theorem}
\begin{proof}
To prove the result, we  need to verify both the conditions of Theorem \ref{thmainresult}. For Condition (1) of Theorem \ref{thmainresult}, let $\mu\in\mathbb{F}_2^{m*}$ and let $\nu,\nu'\in\mathbb{F}_2^n$ with $\nu\neq \nu'$, by Proposition \ref{propwalshtransform} one gets
\begin{align*}
W_F(\mu,\nu)\pm W_F(\mu,\nu')=
\begin{cases}W_G(\tilde{\mu},\nu)\pm W_G(\tilde{\mu},\nu'), &{\rm~if~}\mu=(0,\tilde{\mu}), \\
W_f(\nu)\pm W_f(\nu'), &{\rm~if~} \mu=(1,\mathbf{0_{m-1}}),\\
W_{A_{\tilde{\mu}}}(\nu)\pm W_{A_{\tilde{\mu}}}(\nu'), &{\rm~if~}\mu=(1, \tilde{\mu}),\tilde{\mu}\neq\mathbf{0_{m-1}}.
\end{cases}
\end{align*}
In each case one always has $W_F(\mu,\nu)\pm W_F(\mu,\nu')\neq 2^n$. In fact, the first case is because that $C_G$ is minimal, the second case is due to Condition ({\bf a}) of Construction \ref{ourcon2}, while the third case is the same as Condition (1) of this theorem.

Now we verify the second condition of Theorem \ref{thmainresult}. According to the relations $\mu=(\mu_1, \tilde{\mu})$ and $\mu'=(\mu_1', \tilde{\mu}')$, where $\mu, \mu', \mu+\mu'\in\mathbb{F}_2^{m*}$, the discussions can be divided into 10 cases, which are listed in Table \ref{b5}. Since the discussions of some cases are similar, without loss of generality, here we only discuss the cases 1, 2 and 4 in detail.

{\bf Case 1.} $\mu_1=0,\tilde{\mu}\neq\mathbf{0_{m-1}}$, $\mu_1'=0,\tilde{\mu}'\neq\mathbf{0_{m-1}}$, $\tilde{\mu}\neq \tilde{\mu}'$. In this case, by Proposition \ref{propwalshtransform} we have
\begin{align*}
W_F(\mu,\nu)-W_F(\mu',\nu')+W_F(\mu+\mu',\nu+\nu')=W_G(\tilde{\mu},\nu)-W_G(\tilde{\mu}',\nu')+W_G(\tilde{\mu}+\tilde{\mu}',\nu+\nu'),
\end{align*}
which can never be equal to $2^n$ for any $\nu,\nu'\in\mathbb{F}_2^n$, as $C_G$ is minimal.

{\bf Case 2.} $\mu_1=0,\tilde{\mu}\neq\mathbf{0_{m-1}}$, $\mu_1'=1,\tilde{\mu}'=\mathbf{0_{m-1}}$. In this case, by Proposition \ref{propwalshtransform} we have
\begin{align*}
W_F(\mu,\nu)-W_F(\mu',\nu')+W_F(\mu+\mu',\nu+\nu')=W_G(\tilde{\mu},\nu)-W_f(\nu')+W_{A_{\tilde{\mu}}}(\nu+\nu'),
\end{align*}
which does not equal $2^n$ for any $\nu,\nu'\in\mathbb{F}_2^n$ if and only if
\begin{align*}
-W_f(\nu)+W_G(\tilde{\mu},\nu')+W_{A_{\tilde{\mu}}}(\nu+\nu')\neq 2^n
\end{align*}
for any $\nu,\nu'\in\mathbb{F}_2^n$. This corresponds to Condition (2) of this theorem with $i=1$  and  $j=k=0$.

{\bf Case 4.} $\mu_1=0,\tilde{\mu}\neq\mathbf{0_{m-1}}$, $\mu_1'=1,\tilde{\mu}'\neq\mathbf{0_{m-1}}$, $\tilde{\mu}\neq\tilde{\mu}'$. In this case, one has
\begin{align*}
W_F(\mu,\nu)-W_F(\mu',\nu')+W_F(\mu+\mu',\nu+\nu')
=W_G(\tilde{\mu},\nu)-W_{A_{\tilde{\mu}'}}(\nu')+W_{A_{\tilde{\mu}+\tilde{\mu}'}}(\nu+\nu'),
\end{align*}
which does not equal $2^n$ for any $\nu,\nu'\in\mathbb{F}_2^n$ if and only Condition (3) of this theorem holds with $i=0$ and $j=1$, since
in this case the roles of ${\tilde{\mu}}$, ${\tilde{\mu}}'$ and $\tilde{\mu}+\tilde{\mu}'$ are the same, and the roles of $\nu$, $\nu'$ and $\nu+\nu'$ are the same, respectively.

Similarly, it is checked that Case 9 corresponds to Condition (2) of this theorem with $i=1,j=0$  and  $k=0$; Cases 5 and 7 correspond to Condition (2) with $i=0,j=1$  and  $k=0$; Cases 3 and 6 correspond to Condition (2) with $i=0,j=0$  and  $k=1$; and Cases 8 and 10 correspond to Condition (3) with $i=1$  and  $j=0$.

The result then follows from the above discussions.
\end{proof}

\begin{table*}[!htbp]\setlength{\abovecaptionskip}{0cm}
\caption{The cases of $\mu$ and $\mu'$ in the proof of Theorem \ref{thgenericAB}}  \centering \label{b5}
\medskip
\begin{tabular}{|c|c|c|c|c|c|}
 \hline
Cases & $\mu_1$ & $\tilde{\mu}$ & $\mu_1'$ & $\tilde{\mu}'$ &$\tilde{\mu} +\tilde{\mu}'$ \\
 \hline
1 & $0 $ & $\neq\mathbf{0_{m-1}}$ & $0 $ & $\neq\mathbf{0_{m-1}}$ &$\neq\mathbf{0_{m-1}}$ \\
 \hline
 2 & $0 $ & $\neq\mathbf{0_{m-1}}$ & $1 $ & $\mathbf{0_{m-1}}$ &$\neq\mathbf{0_{m-1}}$ \\
 \hline
 3 & $0 $ & $\neq\mathbf{0_{m-1}}$ & $1 $ & $\neq\mathbf{0_{m-1}}$ &$\mathbf{0_{m-1}}$ \\
 \hline
 4 & $0 $ & $\neq\mathbf{0_{m-1}}$ & $1 $ & $\neq\mathbf{0_{m-1}}$ &$\neq\mathbf{0_{m-1}}$ \\
 \hline
 5 & $1 $ & $ \mathbf{0_{m-1}}$ & $0 $ & $\neq\mathbf{0_{m-1}}$ &$\neq\mathbf{0_{m-1}}$ \\
 \hline
6 & $1 $ & $ \mathbf{0_{m-1}}$ & $1 $ & $\neq\mathbf{0_{m-1}}$ &$\neq\mathbf{0_{m-1}}$ \\
 \hline
 7 & $1 $ & $\neq\mathbf{0_{m-1}}$ & $0 $ & $\neq\mathbf{0_{m-1}}$ &$\mathbf{0_{m-1}}$ \\
 \hline
 8 & $1 $ & $\neq\mathbf{0_{m-1}}$ & $0 $ & $\neq\mathbf{0_{m-1}}$ &$\neq\mathbf{0_{m-1}}$ \\
 \hline
 9 & $1 $ & $\neq\mathbf{0_{m-1}}$ & $1 $ & $\mathbf{0_{m-1}}$ &$\neq\mathbf{0_{m-1}}$ \\
 \hline
 10 & $1 $ & $\neq\mathbf{0_{m-1}}$ & $1 $ & $\neq\mathbf{0_{m-1}}$ &$\neq\mathbf{0_{m-1}}$ \\
 \hline
\end{tabular}
\end{table*}

\begin{remark}\label{rmk-m=2}
Note that for $m=2$, all the three conditions of Theorem \ref{thgenericAB} are fulfilled if and only if the first two of them are fulfilled.
\end{remark}

\begin{remark}\label{rmkab}
To find minimal linear codes from Construction \ref{ourcon2}, firstly one has to find a vectorial Boolean function $G$ such that $C_G$ is minimal, and then to find a Boolean function $f$ satisfying  Conditions ${\rm({\bf a})}$ and  ${\rm({\bf b})}$ (In fact, according to \cite[Theorem 15]{Ding-2018-IT} and Remark \ref{rmkabcondition}, Conditions ${\rm({\bf a})}$ and  ${\rm({\bf b})}$ just mean that the linear code $C_f$ in \eqref{eqcodeBooleanintro} is  minimal violating the AB condition). Finally, one has to verify the three conditions of Theorem \ref{thgenericAB}. Therefore, compared with the construction of minimal linear codes violating the AB condition from Boolean functions, the method of this paper by using vectorial Boolean functions is more challenging.
\end{remark}

\section{Several minimal linear codes violating the AB condition}\label{sec:violating}

In this section, let $n=2t$. Let $E_0,E_1,\ldots, E_{2^t}$ be $t$-dimensional linear subspaces of $\mathbb{F}_2^n$ such that $E_i\cap E_j=\{\mathbf{0_n}\}$ for any $0\leq i< j \leq 2^t$. Let $1_{E_i}(x)$ be the  characteristic function of $E_i$, which is equal to 1 if $x\in E_i$, and is equal to 0 otherwise.
Denote by  $k=(k_0,k_1,\ldots, k_{t-1})_2$ the binary expansion of the integer $0\leq k\leq 2^t-1$, this is, $k=\sum_{i=0}^{t-1}k_i2^i$. This representation is also regarded as a vector of $\mathbb{F}_2^t$ for convenience.  We give several minimal linear codes violating the AB condition by using Theorem \ref{thgenericAB}.
In the proof of our main result, we need the following lemma, which is inspired by \cite[Lemma 16]{Ding-2018-IT}.

\begin{lemma}\label{lemding}
Let $f(x)=\sum_{i=1}^s1_{E_i}(x)$ for some $1\leq s\leq 2^t$. Then the Walsh-Hadamard transform of $f$ is given by
\begin{align*}
W_f(\nu)=\begin{cases}
2^n-2s(2^t-1)-2f(\mathbf{0_n}),\hspace{0.3cm}&{\rm if}~\nu={\mathbf{0_n}};\\
-2^{t+1}+2s-2f(\mathbf{0_n}), \hspace{0.3cm}&{\rm if}~\nu\in \cup_{i=1}^s E_i^{\bot}\setminus \{\mathbf{0_n}\};\\
2s-2f(\mathbf{0_n}), \hspace{0.3cm}&{\rm otherwise}.
\end{cases}
\end{align*}
\end{lemma}
\begin{proof}
By the relation $(-1)^{f(x)}=1-2f(x)$, for any $\nu\in\mathbb{F}_2^n$, we have
\begin{align*}
W_f(\nu)=&\sum_{x\in\mathbb{F}_2^n}(-1)^{\nu\cdot x}-2\sum_{x\in\mathbb{F}_2^n}f(x)(-1)^{\nu\cdot x}\\
=& 2^n\delta_0(\nu)-2\bigg(f(\mathbf{0_n})+\sum_{i=1}^s\bigg(\sum_{x\in E_i}(-1)^{\nu\cdot x}-1\bigg)\bigg),
\end{align*}
where $\delta_0(x)$ equals 1 if $x={\bf0_n}$, and equals 0 otherwise. Then the result follows from the facts that $\sum_{x\in E_i}(-1)^{\nu\cdot x}=2^t1_{E_i^{\bot}}(\nu)$ with $E_i^{\bot}=\{x\in\mathbb{F}_2^n : \nu\cdot x=0, ~\forall ~\nu\in E_i\}$, and that $E_i^{\bot}\cap E_j^{\bot}=\{\mathbf{0_n}\}$ for any $1\leq i< j \leq 2^t$.
\end{proof}

\subsection{The case $n$ even and $m=2$}

In this subsection, we present an application of Theorem \ref{thgenericAB} for the case $n$ even and $m=2$, that is, for the case $F=(f,g)$, where $f$ and $g$ are two Boolean functions on $\mathbb{F}_2^n$.

\begin{theorem}\label{thfirstvioab}
Let $n=2t$ with $t\geq3$. Let $f(x)=1_{E_0}(x)+1_{E_{2^t}}(x)$ and
  \begin{align}\label{eqps}
g_i(x)=\sum_{0\leq k=(k_0,k_1,\ldots,k_{t-1})_2\leq 2^t-1, k_i=1}1_{E_k}(x)+1_{E_{2^t}}(x)+1, ~\forall~0\leq i\leq t-1.
\end{align}
  Let $F_i$ be the $(n, 2)$-function given by $F_i(x)=(f(x), g_i(x))$,  and let $C_{F_i}$ be the linear code generated by Construction \ref{ourcon}. Then $C_{F_i}$ is a minimal linear code with parameters $[2^n-1,n+2, 2^{t+1}-2]$  and weight distribution in Table \ref{b6}. Moreover, $C_{F_i}$ violates the AB condition.
\end{theorem}
\begin{proof}
By Lemma \ref{lemding}, we have
\begin{align}\label{eqwalsh-f}
W_f(\nu)=\begin{cases}
2^n-4(2^t-1),\hspace{0.3cm}&{\rm if}~\nu={\mathbf{0_n}};\\
-2^{t+1}+4, \hspace{0.3cm}&{\rm if}~\nu\in (E_0^{\bot}\cup E_{2^{t}}^{\bot})\setminus\{\mathbf{0_n}\};\\
4, \hspace{0.3cm}&{\rm otherwise},
\end{cases}
\end{align}
which clearly satisfies Conditions ({\bf a}) and ({\bf b}) in Construction \ref{ourcon2} when $t\geq 2$. For any $0\leq i\leq t-1$, we also have
\begin{align}\label{eqWalsh-g-1}
W_{g_i}(\nu)=\begin{cases}
2^{t}, \hspace{0.3cm}&{\rm if}~\nu\in (\bigcup_{0\leq k\leq 2^{t}-1, k_i=1}E_k^{\bot})\cup E_{2^t}^{\bot};\\
-2^t,\hspace{0.3cm}&{\rm otherwise,}
\end{cases}
\end{align}
 which is a bent function with $g_i(\mathbf{0_n})=0$. Hence $C_{g_i}$ is a minimal linear code by Corollary \ref{corVB}. Consequently, $f$ and every $g_i$ satisfy the conditions of Construction \ref{ourcon2}.
 Therefore, each $C_{F_i}$ violates the AB condition by Proposition \ref{propvoiAB}.

  Then by Remark \ref{rmk-m=2}, to show $C_{F_i}$ is minimal for any $0\leq i\leq t-1$, we need to verify the first two conditions of Theorem \ref{thgenericAB}. It is noted that
  \begin{align*}
  f(x)+g_i(x)=\sum_{0\leq k=(k_0,k_1,\ldots,k_{t-1})_2\leq 2^t-1, k_i=1}1_{E_k}(x)+1_{E_{0}}(x)+1.
  \end{align*}
 Let $A_i(x)=f(x)+g_i(x)$. For any $\nu\in\mathbb{F}_2^n$, it holds
\begin{align}\label{eqWalsh-A}
W_{A_i}(\nu)
=\begin{cases}
2^{t}, \hspace{0.3cm}&{\rm if}~\nu\in (\bigcup_{0\leq k\leq 2^{t}-1, k_i=1}E_k^{\bot})\cup E_{0}^{\bot};\\
-2^t,\hspace{0.3cm}&{\rm otherwise.}
\end{cases}
\end{align}
Hence, for any $\nu,\nu'\in\mathbb{F}_2^n$ with $\nu\neq \nu'$, we have
\begin{align*}
W_{A_i}(\nu)\pm W_{A_i}(\nu')\in\{0, \pm2^{t+1}\},
\end{align*}
which can never be equal to $2^n$ when $t\geq2$. In other words, Condition (1) of  Theorem \ref{thgenericAB} is satisfied.
 In addition, for any $\nu,\nu'\in\mathbb{F}_2^n$ and any $p,q,r\in\mathbb{F}_2$, it is easily checked that
 \begin{align*}
(-1)^pW_f(\nu)+(-1)^qW_{g_i}(\nu')+(-1)^{r}W_{A_i}(\nu+\nu')\in\bigg\{\pm (2^n-4(2^t-1)),\pm(2^n-2^{t+1}+4),\\
 \pm(2^n-6\times 2^t+4), \pm(2^{t+1}-4), \pm(2^{t+1}+4), \pm(2^{t+2}-4), \pm4\bigg\},
\end{align*}
 which also can never be equal to $2^n$ when $t\geq 2$, and thus Condition (2) of Theorem \ref{thgenericAB} is satisfied. Hence $C_{F_i}$ is minimal. Moreover, according to Propositions \ref{propweightdistri} and \ref{propwalshtransform}, and Relations \eqref{eqwalsh-f}, \eqref{eqWalsh-g-1} and \eqref{eqWalsh-A}, we can obtain the weight distribution of $C_{F_i}$, as listed in Table \ref{b6}.

 \begin{table*}[!htbp]\setlength{\abovecaptionskip}{0cm}
\caption{Weight distribution of $C_{F_i}$ in Theorem \ref{thfirstvioab}}  \centering \label{b6}
\medskip
\begin{tabular}{|c|c|}
 \hline
 Weight & Frequency\\
 \hline
 0 & 1\\
 \hline
 $2^{n-1}$ & $2^{n}-1$\\
  \hline
 $2^{t+1}-2$ & $1$\\
  \hline
 $2^{n-1}+2^{t}-2$ & $2^{t+1}-2$\\
 \hline
 $2^{n-1}-2$ & $2^n-2^{t+1}+1$\\
  \hline
 $2^{n-1}-2^{t-1}$ & $2^n+2^t$\\
 \hline
$2^{n-1}+2^{t-1}$ & $2^n-2^t$\\
 \hline
\end{tabular}
\end{table*}
\end{proof}

\begin{remark}\label{rmkk=0}
 Let $f(x)=1_{E_{2^t-1}}(x)+1_{E_{2^t}}(x)$, and
  \begin{align}\label{eqpsk=0}
g_i(x)=\sum_{0\leq k=(k_0,k_1,\ldots,k_{t-1})_2\leq 2^t-1, k_i=0}1_{E_k}(x)+1_{E_{2^t}}(x)+1, ~\forall ~0\leq i\leq t-1.
\end{align}
Then one can get a similar result as that of Theorem \ref{thfirstvioab}.
\end{remark}

\begin{example}\label{exa3}
Let $n=6$. Then the linear code $C_{F_i}$ in Theorem \ref{thfirstvioab} is a minimal binary linear code with parameters $[63, 8, 14]$ and  weight enumerator
\begin{align*}
1+z^{14}+72z^{28}+49z^{30}+63z^{32}+56z^{36}+14z^{38}.
\end{align*}
Clearly, $w_{\min}/w_{\max}=14/38<1/2$.
\end{example}

Comparing the parameters of $C_{F_i}$ in Example \ref{exa3} with \cite[Example 20]{Ding-2018-IT}, the dimension of $C_{F_i}$ is larger than that of
\cite[Example 20]{Ding-2018-IT}.

It is noted that Theorem \ref{thfirstvioab} cannot be generalized to the case $m>2$ by letting $G=(g_0, g_1,\ldots,g_{r-1})$ with $2\leq r\leq t$ (which is a vectorial bent function by Lemma \ref{thli-ipl}), even the Boolean function $f$ in Theorem \ref{thfirstvioab} and the newly defined function $G$ satisfy the conditions of Construction \ref{ourcon2}. Because in this case, one can check that though the conditions of Theorem \ref{thgenericAB} are satisfied  when ${\rm wt}(\tilde{\mu})$ is odd, Condition (2) of Theorem \ref{thgenericAB} is not satisfied  when ${\rm wt}(\tilde{\mu})$ is even. So we have to find another function $f$ satisfying the conditions of Construction \ref{ourcon2} for generalization, which is done in the next section.

\subsection{The case $n$ even and $2< m\leq t+1$}

In this subsection, we present an application of Theorem \ref{thgenericAB} for the case $n$ even and $2< m\leq t+1$, that is, the case $F=(f, G)$, where $f$ is a Boolean function on $\mathbb{F}_2^n$, and $G$ is an $(n,r)$-function with $2\leq r=m-1\leq t$. To this end, we first give a class of vectorial bent functions, which is obtained by modifying the vectorial bent functions in \cite{Li-IPL,Li-IET}.

\begin{lemma}\label{thli-ipl}
Let $G=(g_0,g_1,\ldots,g_{r-1})$ be an $(n,r)$-function with $2\leq r\leq t$, $g_i$ is defined by \eqref{eqps} for any $0\leq i\leq r-1$.
Then $G$ is a vectorial bent function.
\end{lemma}
\begin{proof}
For any $\tilde{\mu}=(\mu_0,\mu_1,\ldots,\mu_{r-1})\in\mathbb{F}_2^{r*}$, we have
\begin{align}\label{eqmuG}
\tilde{\mu}\cdot G(x)
=&\sum_{0\leq k\leq 2^t-1}\bigg(\sum_{0\leq i\leq r-1,k_i=1}\mu_i \bigg)1_{E_k}(x)+(1_{E_{2^t}}(x)+1)\sum_{i=0}^{r-1}\mu_i\nonumber\\
=&\sum_{0\leq k\leq 2^t-1, \tilde{\mu}\cdot \tilde{k}=1}1_{E_k}(x)+(1_{E_{2^t}}(x)+1)\sum_{i=0}^{r-1}\mu_i\nonumber\\
=&\begin{cases}
\sum_{0\leq k\leq 2^t-1, \tilde{\mu}\cdot \tilde{k}=1}1_{E_k}(x)+1_{E_{2^t}}(x)+1,\hspace{0.3cm} &{\rm if}~{\rm wt}(\tilde{\mu})~{\rm is~odd};\\
 \sum_{0\leq k\leq 2^t-1, \tilde{\mu}\cdot \tilde{k}=1}1_{E_k}(x),\hspace{0.3cm} &{\rm if}~{\rm wt}(\tilde{\mu})~{\rm is~even},
\end{cases}
\end{align}
where $\tilde{k}=(k_0,k_1,\ldots, k_{r-1})$.   Since the cardinality of the set
 $$\bigg\{0\leq k=(k_0,k_1,\ldots,k_{t-1})_2\leq 2^t-1 : \tilde{\mu}\cdot \tilde{k}=1, \tilde{\mu}\in\mathbb{F}_2^{r*}\bigg\}$$
 is equal to $2^{t-1}$, by Lemma \ref{lemding} we have:
\begin{enumerate}
\item[1)] When ${\rm wt}(\tilde{\mu})$ is odd, it holds that
\begin{align}\label{eqWalsh-G-odd}
W_G(\tilde{\mu},\nu)=\begin{cases}
2^{t}, \hspace{0.3cm}&{\rm if}~\nu\in (\bigcup_{0\leq k\leq 2^{t}-1, \tilde{\mu}\cdot \tilde{k}=1}E_k^{\bot})\cup E_{2^t}^{\bot};\\
-2^t,\hspace{0.3cm}&{\rm otherwise.}
\end{cases}
\end{align}
\item[2)] When ${\rm wt}(\tilde{\mu})$ is even, it holds that
\begin{align}\label{eqWalsh-G-even}
W_G(\tilde{\mu},\nu)=\begin{cases}
-2^{t}, \hspace{0.3cm}&{\rm if}~\nu\neq\mathbf{0_n},\nu\in \bigcup_{0\leq k\leq 2^{t}-1,\tilde{\mu}\cdot \tilde{k}=1}E_k^{\bot};\\
2^t,\hspace{0.3cm}&{\rm otherwise.}
\end{cases}
\end{align}
\end{enumerate}
This completes the proof.
\end{proof}

To give an infinite family of minimal binary linear codes violating the AB condition, we also need the following results.

\begin{lemma}\cite{Carlet-2006,Sihem-2014}\label{lemcarlet}
Let $\varphi_1,\varphi_2$ and $\varphi_3$ be three Boolean functions on $\mathbb{F}_2^n$. Let $f(x)=\varphi_1(x)\varphi_2(x)+\varphi_1(x)\varphi_3(x)+\varphi_2(x)\varphi_3(x)$. Then for any $\nu\in\mathbb{F}_2^n$, it holds that
\begin{align*}
W_f(\nu)=\frac{1}{2}\big(W_{\varphi_1}(\nu)+W_{\varphi_2}(\nu)+W_{\varphi_3}(\nu)-W_{\varphi_4}(\nu)\big),
\end{align*}
where $\varphi_4=\varphi_1+\varphi_2+\varphi_3$.
\end{lemma}

A generalized version of Lemma \ref{lemcarlet} is referred to \cite{Li et al.-2021}.

\begin{lemma}\label{lem-f}
Let $n=2t$ with $t\geq 2$. Let $a,b\in\mathbb{F}_2^{n*}$ with $a\neq b$, and let $E$ be a $t$-dimensional linear subspace of $\mathbb{F}_2^n$. Then the Boolean function $f$ on $\mathbb{F}_2^n$ defined by
\begin{align}\label{eq-f}
f(x)=1_E(x)+(a\cdot x)(b\cdot x)+1
\end{align}
satisfies Condition {\rm({\bf a})} of Construction \ref{ourcon2} if and only if $a,b,a+b\notin E^{\bot}$.
\end{lemma}
\begin{proof}
Let $\varphi(x)=1_E(x)+1$. By Lemma \ref{lemding}, one has
\begin{align}\label{eqwalsh-1E}
W_{\varphi}(\nu)=\begin{cases}
-2^n+2^{t+1},\hspace{0.3cm}&{\rm if}~\nu={\bf 0_n};\\
2^{t+1},\hspace{0.3cm}&{\rm if}~\nu\in E^{\bot}\backslash\{\bf 0_n\};\\
0,\hspace{0.3cm}&{\rm otherwise}.
\end{cases}
\end{align}
Note that $f(x)=\varphi(x)+(a\cdot x)(b\cdot x)=\varphi_1(x)\varphi_2(x)+\varphi_1(x)\varphi_3(x)+\varphi_2(x)\varphi_3(x)$, where $\varphi_1(x)=\varphi(x)$, $\varphi_2(x)=\varphi(x)+a\cdot x$ and $\varphi_3(x)=\varphi(x)+b\cdot x$. By Lemma \ref{lemcarlet}, we have
\begin{align}\label{eqwalsh-f-1E}
W_f(\nu)=\frac{1}{2}\big(W_{\varphi}(\nu)+W_{\varphi}(\nu+a)+W_{\varphi}(\nu+b)-W_{\varphi}(\nu+a+b)\big),~\forall~\nu\in\mathbb{F}_2^n.
\end{align}

Now if $a,b,a+b\notin E^{\bot}$,  we can deduce from \eqref{eqwalsh-f-1E} that
\begin{align}\label{eqwalsh-f-1E-2}
W_f(\nu)=\begin{cases}
-2^{n-1}+2^t, \hspace{0.3cm}&{\rm if}~\nu\in\{{\bf 0_n}, a, b\};\\
2^{n-1}-2^t, \hspace{0.3cm}&{\rm if}~\nu=a+b;\\
2^t,\hspace{0.3cm}&{\rm if}~\nu \in (E^{\bot}\cup a+E^{\bot}\cup b+E^{\bot})\setminus\{{\bf 0_n}, a, b, a+b\};\\
-2^t,\hspace{0.3cm}&{\rm if}~\nu\in (a+ b+E^{\bot})\setminus\{{\bf 0_n}, a, b, a+b\};\\
0,\hspace{0.3cm}&{\rm otherwise}.
\end{cases}
\end{align}
Then it is easily checked that Condition {({\bf a})} of Construction \ref{ourcon2} is satisfied when $t\geq 2$.

Conversely, assume that $a,b\in E^{\bot}\backslash\{{\bf 0_n}\}$ with $a\neq b$, we deduce from \eqref{eqwalsh-f-1E} that
\begin{align*}
W_f({\bf 0_n})=W_f(a)=W_f(b)=-2^{n-1}+2^{t+1} {\hspace{0.3cm}}{\rm and}{\hspace{0.3cm}}W_f(a+b)=2^{n-1}+2^{t+1},
\end{align*}
from which we can get $W_f(a+b)-W_f(a)=2^n$. This means that Condition {({\bf a})} of Construction \ref{ourcon2} is not satisfied for the case $a,b\in E^{\bot}\backslash\{{\bf 0_n}\}$ with $a\neq b$. Similarly, one can show that  Condition {({\bf a})} is also not satisfied for the cases ${\bf 0}\neq a\in E^{\bot}, b\notin E^{\bot}$ and ${\bf 0}\neq b\in E^{\bot}, a\notin E^{\bot}$. This completes the proof.
\end{proof}

\begin{remark}\label{rmk-f(b)}
When $t\geq 3$, it is easily checked that the function $f$ in Lemma \ref{lem-f} also satisfies Condition {\rm({\bf b})} of Construction \ref{ourcon2}.
\end{remark}

\begin{lemma}\label{5-valued}
Let $\varphi$ be a bent function on $\mathbb{F}_{2}^{2t}$. Let $a,b\in\mathbb{F}_2^{2t}$, and $h(x)=\varphi(x)+(a\cdot x)(b\cdot x)$. Then for any $\nu\in\mathbb{F}_2^{2t}$, it holds that $W_{h}(\nu)\in\{0,\pm 2^t, \pm 2^{t+1}\}.$
\end{lemma}
\begin{proof}
Since $\varphi$ is bent, by \eqref{eqwalsh-f-1E}, for any $\nu\in\mathbb{F}_2^{2t}$, we have
\begin{align*}
W_h(\nu)=2^{t-1}\big((-1)^{\varphi^*(\nu)}+(-1)^{\varphi^*(\nu+a)}+(-1)^{\varphi^*(\nu+b)}-(-1)^{\varphi^*(\nu+a+b)}\big),
\end{align*}
where $\varphi^*$ is the dual of $\varphi$. Note that
\begin{align*}
(-1)^{\varphi^*(\nu)}+(-1)^{\varphi^*(\nu+a)}+(-1)^{\varphi^*(\nu+b)}-(-1)^{\varphi^*(\nu+a+b)}\in\{0,\pm2,\pm4\}
 \end{align*}
 for any $a,b\in\mathbb{F}_2^{2t}$. Therefore,  $W_{h}(\nu)\in\{0,\pm 2^t, \pm 2^{t+1}\}$ for any $\nu\in\mathbb{F}_2^{2t}$.
\end{proof}

Now we give an infinite family of minimal binary linear codes violating the AB condition.

\begin{theorem}\label{th-2<m<t}
Let $n=2t$ with $t\geq 3$. Let $f$ be a Boolean function on $\mathbb{F}_2^n$ defined by
\begin{align*}
f(x)=1_{E^{2^t}}(x)+(a\cdot x)(b\cdot x)+1,
\end{align*}
where $a,b,a+b\in E_0^{\bot}\backslash\{{\bf0_n}\}$. Let $G$ be the $(n,r)$-bent function generated by Lemma \ref{thli-ipl} with $2\leq r\leq t$. Let $F=(f,G)$, and let $C_F$ be the binary linear code generated by Construction \ref{ourcon}. Then $C_F$ is a minimal linear code with parameters $[2^n-1, n+r+1,2^{n-2}+2^{t-1}]$ violating the AB condition.
\end{theorem}
\begin{proof}
Since $a,b,a+b\in E_0^{\bot}\backslash\{{\bf0_n}\}$, one has $a, b, a+b\notin E_{2^t}^{\bot}\backslash\{{\bf0_n}\}$. By Lemma \ref{lem-f} and Remark \ref{rmk-f(b)}, we obtain that $f$ is a Boolean function satisfying Conditions $({\bf a})$ and $({\bf b})$ of Construction \ref{ourcon2}. In addition, as $G$ is an $(n,r)$-bent function with $G({\bf0_n})={\bf0_r}$, $C_G$ defined by \eqref{conG} is a minimal linear code by Corollary \ref{corVB}. Consequently, $f$ and $G$ satisfy the conditions of Construction \ref{ourcon2}, and thus $C_F$ violates the AB condition by Proposition \ref{propvoiAB}. Below we show that $C_F$ is minimal, for which we need to verify the three conditions of Theorem \ref{thgenericAB}.

Firstly, for any $\tilde{\mu}\in\mathbb{F}_2^{r*}$,  we obtain from \eqref{eqmuG} that
\begin{align*}
f(x)+\tilde{\mu}\cdot G(x)=\begin{cases}
\sum_{0\leq k\leq 2^t-1, \tilde{\mu}\cdot \tilde{k}=1}1_{E_k}(x)+(a\cdot x)(b\cdot x),\hspace{0.3cm} &{\rm if}~{\rm wt}(\tilde{\mu})~{\rm is~odd};\\
 \sum_{0\leq k\leq 2^t-1, \tilde{\mu}\cdot \tilde{k}=1}1_{E_k}(x)+1_{E_{2^t}}(x)+(a\cdot x)(b\cdot x)+1,\hspace{0.3cm} &{\rm if}~{\rm wt}(\tilde{\mu})~{\rm is~even}.
\end{cases}
\end{align*}
Let $A_{\tilde{\mu}}(x)=f(x)+\tilde{\mu}\cdot G(x)$. Then by \eqref{eqmuG}, \eqref{eqWalsh-G-odd}, \eqref{eqWalsh-G-even} and Lemma \ref{5-valued},  we obtain that
\begin{align}\label{eq-Amu}
W_{A_{\tilde{\mu}}}(\nu)
\in \{0, \pm2^t, \pm2^{t+1}\}, ~~\forall~\nu\in\mathbb{F}_2^n.
\end{align}
 Hence, for any $\tilde{\mu}\in\mathbb{F}_2^{r^*}$, and any $\nu,\nu'\in\mathbb{F}_2^n$ with $\nu\neq\nu'$, we have
 \begin{align*}
 W_{A_{\tilde{\mu}}}(\nu)\pm  W_{A_{\tilde{\mu}}}(\nu')\in\{0,\pm 2^t,\pm2^{t+1},\pm 3\cdot 2^t, \pm 2^{t+2}\},
 \end{align*}
 which can never be equal to $2^n$ when $t\geq 3$. So Condition (1)  of  Theorem \ref{thgenericAB} is satisfied. In addition, for any $\tilde{\mu}\in\mathbb{F}_2^{r*}$, $\nu,\nu'\in\mathbb{F}_2^n$, and for any $i,j,k\in\mathbb{F}_2$, it is easy to check from \eqref{eqWalsh-G-odd}, \eqref{eqWalsh-G-even}, \eqref{eqwalsh-f-1E-2} and \eqref{eq-Amu} that
 \begin{align*}
 (-1)^iW_f(\nu)&+(-1)^jW_G(\tilde{\mu},\nu')+(-1)^kW_{A_{\tilde{\mu}}}(\nu+\nu')\in\bigg\{0,\pm2^t,\pm2^{t+1},\pm3\cdot2^{t},\pm2^{t+2}, \pm2^{n-1},\\
 &\pm2^{n-1}\pm2^t, \pm2^{n-1}\pm2^{t+1},\pm(2^{n-1}-2^{t+1}),\pm(2^{n-1}-3\cdot2^{t}),\pm(2^{n-1}-2^{t+2})\bigg\},
 \end{align*}
which  also cannot be equal to $2^n$ when $t>2$. Hence Condition (2)  of  Theorem \ref{thgenericAB} is satisfied. Moreover, for any $\tilde{\mu}, \tilde{\mu}'\in\mathbb{F}_2^{r*}$ with $\tilde{\mu}\neq\tilde{\mu}'$, any $\nu,\nu'\in\mathbb{F}_2^n$, and for any $i, j\in\mathbb{F}_2$, it is easy to check from \eqref{eqWalsh-G-odd}, \eqref{eqWalsh-G-even} and \eqref{eq-Amu} that
\begin{align*}
W_{A_{\tilde{\mu}}}(\nu)+(-1)^iW_G(\tilde{\mu}',\nu')+(-1)^jW_{A_{\tilde{\mu}+\tilde{\mu}'}}(\nu+\nu')\in\big\{0,\pm2^t,\pm2^{t+1},\pm3\cdot2^{t},
\pm2^{t+2},\pm5\cdot2^t\big\},
\end{align*}
which can never be equal to $2^n$ when $t>2$. Therefore, Condition (3)  of  Theorem \ref{thgenericAB} is also satisfied. Consequently, $C_F$ is minimal by Theorem \ref{thgenericAB}.

Finally, according to Propositions \ref{propdim} and \ref{propwalshtransform}, Remark \ref{rmkdistance}, Lemma \ref{thli-ipl}, and Equations \eqref{eqwalsh-f-1E-2} and \eqref{eq-Amu}, one can obtain that $C_F$ is a $[2^n-1,n+r+1,2^{n-2}+2^{t-1}]$ code. The proof is finished.
\end{proof}

Note that the weight distribution of $C_F$ in Theorem \ref{th-2<m<t} is not obtained. It seems difficult to get the exact numbers of $\nu\in\mathbb{F}_2^n$ in \eqref{eq-Amu} such that $W_{A_{\tilde{\mu}}}(\nu)=0, 2^t,-2^t,2^{t+1},-2^{t+1}$, respectively. If those numbers are obtained, then by Propositions \ref{propweightdistri} and \ref{propwalshtransform}, Relations \eqref{eqWalsh-G-odd}, \eqref{eqWalsh-G-even} and \eqref{eqwalsh-f-1E-2}, the weight distribution of $C_F$ can also be obtained.

\begin{remark}\label{rmk-k=0m>2}
In Lemma \ref{thli-ipl}, let $g_i$ be defined by \eqref{eqpsk=0} for any $0\leq i\leq r-1$. Then for any $\tilde{\mu}\in \mathbb{F}_2^{r*}$, one has
\begin{align*}
\tilde{\mu}\cdot G(x)=\begin{cases}
\sum_{0\leq k\leq 2^t-1, \tilde{\mu}\cdot \tilde{k}=0}1_{E_k}(x)+1_{E_{2^t}}(x)+1,\hspace{0.3cm} &{\rm if}~{\rm wt}(\tilde{\mu})~{\rm is~odd};\\
 \sum_{0\leq k\leq 2^t-1, \tilde{\mu}\cdot \tilde{k}=1}1_{E_k}(x),\hspace{0.3cm} &{\rm if}~{\rm wt}(\tilde{\mu})~{\rm is~even},
\end{cases}
\end{align*}
which also implies that $G=(g_0,g_1,\ldots,g_{r-1})$ is vectorial bent. Applying this result to Theorem \ref{th-2<m<t}, one can obtain a similar result as that of Theorem \ref{th-2<m<t}.
\end{remark}

\subsection{The case $n$ even and $m=n+1$}

In this subsection, let $n=2t$, $i$ be an positive integer and $\lambda=\gcd(n,i)$ such that $\frac{n}{\lambda}$ is odd. We present an application of Theorem \ref{thgenericAB} for the case $m=n+1$, that is, the case $F=(f,G)$, where $f$ is a Boolean function on $\mathbb{F}_{2^n}$ and $G$ is an $(n, n)$-function. Here we choose $G(x)=x^{2^i+1}$ to be a Gold function. We need the following known result on  Gold functions.

\begin{lemma}\cite{Gold-1999}\label{Gold-walsh}
Let $\lambda=\gcd(n,i)$ such that $\frac{n}{\lambda}$ is odd. Then $G(x)=x^{2^i+1}$ is a vectorial plateaued $(n,n)$-function with single amplitude $2^{\frac{n+\lambda}{2}}$.
\end{lemma}

We also need the following result.

\begin{lemma}\label{lemf+G}
Let $n=2t$ and $\lambda=\gcd(n,i)$ such that $\frac{n}{\lambda}$ is odd. Let $E=\mathbb{F}_{2^t}$. For any $\tilde{\mu}\in\mathbb{F}_{2^n}^*$, let
\begin{align}\label{eqvarphi}
\varphi(x)={\rm Tr}_1^n(\tilde{\mu}x^{2^i+1})+1_{E}(x).
\end{align}
Then for any $\nu\in\mathbb{F}_{2^n}$, it holds that
$$W_{\varphi}(\nu)\in\bigg\{0,-2^{t+1},\pm2^{\frac{n+\lambda}{2}},\pm2^{\frac{t+\frac{\lambda}{2}}{2}+1},\pm2^{\frac{n+\lambda}{2}}-2^{t+1}, \pm2^{\frac{n+\lambda}{2}}\pm2^{\frac{t+\frac{\lambda}{2}}{2}+1}\bigg\}.$$
\end{lemma}
\begin{proof}
The Walsh-Hadamard transform of $\varphi$ at $\nu\in\mathbb{F}_{2^n}$ is that
\begin{align*}
W_{\varphi}(\nu)=&\sum_{x\in\mathbb{F}_{2^n}}(-1)^{\varphi(x)+{\rm Tr}_1^n(\nu x)}\\
=&\sum_{x\in\mathbb{F}_{2^n}}(-1)^{{\rm Tr}_1^n(\tilde{\mu}x^{2^i+1}+\nu x)}
-2\sum_{x\in \mathbb{F}_{2^t}}(-1)^{{\rm Tr}_1^t\big((\tilde{\mu}+\tilde{\mu}^{2^t})x^{2^i+1}+(\nu+\nu^{2^t})x\big)}.
\end{align*}
Observe that
\begin{align*}
\sum_{x\in \mathbb{F}_{2^t}}(-1)^{{\rm Tr}_1^t\big((\tilde{\mu}+\tilde{\mu}^{2^t})x^{2^i+1}+(\nu+\nu^{2^t})x\big)}=\begin{cases}
2^t,\hspace{0.3cm}&{\rm if}~\tilde{\mu}\in\mathbb{F}_{2^t}^*,\nu\in\mathbb{F}_{2^t};\\
0, \hspace{0.3cm}&{\rm if}~\tilde{\mu}\in\mathbb{F}_{2^t}^*,\nu\notin\mathbb{F}_{2^t}.
\end{cases}
\end{align*}
Then by Lemma \ref{Gold-walsh}, we obtain that
\begin{align*}
\sum_{x\in\mathbb{F}_{2^n}}(-1)^{{\rm Tr}_1^n(\tilde{\mu}x^{2^i+1}+\nu x)}\in\bigg\{0,\pm2^{\frac{n+\lambda}{2}}\bigg\}
\end{align*}
and
\begin{align*}
\sum_{x\in \mathbb{F}_{2^t}}(-1)^{{\rm Tr}_1^t\big((\tilde{\mu}+\tilde{\mu}^{2^t})x^{2^i+1}+(\nu+\nu^{2^t})x\big)}\in\bigg\{0,\pm 2^{\frac{t+\frac{\lambda}{2}}{2}},2^t\bigg\},
\end{align*}
since $\lambda=2\gcd(t, i)$ and $\frac{t}{\gcd(t,i)}=\frac{n}{\lambda}$ as $\frac{n}{\lambda}$ is odd. The result follows then from the above two relations.
\end{proof}

Now we present our main result of this subsection.

\begin{theorem}\label{thGold}
Let $n=2t>8$, $2\leq i\leq n-1$ and $\lambda=\gcd(n,i)$ such that $\frac{n}{\lambda}$ is odd. Let $E=\mathbb{F}_{2^t}$ and $a,b\in\mathbb{F}_{2^n}\backslash E$ such that $a+b\notin E$. Let $G(x)=x^{2^i+1}$, $f(x)=1_E(x)+{\rm Tr}_1^n(ax){\rm Tr}_1^n(bx)+1$ and $F=(f,G)$. Let $C_F$ be the linear code given by \eqref{eqourconfield}. Then $C_F$ is a minimal linear code violating the AB condition with parameters $[2^n-1,2n+1,2^{n-2}+2^{t-1}]$.
\end{theorem}
\begin{proof}
 Note that $E^{\bot}=\{x\in\mathbb{F}_{2^n} : {\rm Tr}_1^{n}(xy)=0,~\forall~y\in E\}=\mathbb{F}_{2^t}=E$. Then by Lemma \ref{lem-f} and Remark \ref{rmk-f(b)}, $f$ is a Boolean function satisfying Conditions $({\bf a})$ and $({\bf b})$ of Construction \ref{ourcon2}. By Lemma \ref{Gold-walsh} and Corollary \ref{corsingleplateaued}, it is also clear that $C_G$ is a minimal linear code. Therefore, $f$ and $G$ satisfy the conditions of Construction \ref{ourcon2}, and hence $C_F$ violates the AB condition by Proposition \ref{propvoiAB}. Below we show that $C_F$ is minimal, for which we need to verify the three conditions of Theorem \ref{thgenericAB}.

 For any $\tilde{\mu}\in\mathbb{F}_{2^n}^*$, let
 \begin{align*}
 A_{\tilde{\mu}}(x)=f(x)+{\rm Tr}_1^n(\tilde{\mu} G(x))=\varphi(x)+{\rm Tr}_1^n(ax){\rm Tr}_1^n(bx)+1,
 \end{align*}
 where $\varphi$ is the function in \eqref{eqvarphi}. Then by Lemmas \ref{lemcarlet} and \ref{lemf+G}, and Relation \eqref{eqwalsh-f-1E}, we have
 \begin{align}\label{maxWA}
 \max_{(\tilde{\mu},\nu)\in \mathbb{F}_{2^n}^*\times\mathbb{F}_{2^n}} |W_{A_{\tilde{\mu}}}(\nu)|\leq 2^{\frac{n+\lambda}{2}+1}+2^{t+2},
 \end{align}
 since $\lambda\leq t$ (recall that $\lambda=\gcd(n,i)$, $2\leq i\leq n-1$). Hence, for any $\tilde{\mu}\in\mathbb{F}_{2^n}^*$ and any $\nu,\nu'\in\mathbb{F}_{2^n}$ with $\nu\neq\nu'$, we have
 \begin{align*}
 W_{A_{\tilde{\mu}}}(\nu)\pm W_{A_{\tilde{\mu}}}(\nu')
\leq 2^{\frac{n+\lambda}{2}+2}+2^{t+3},
 \end{align*}
 which is strictly less than $2^n$ when $t>4$.
 For any $\tilde{\mu}\in\mathbb{F}_{2^n}^*$, any $\nu,\nu'\in\mathbb{F}_{2^n}$ and for any $p,q,r\in\mathbb{F}_2$, by Lemma \ref{Gold-walsh} and Relations \eqref{eqwalsh-f-1E-2} and \eqref{maxWA}, we have
 \begin{align*}
 (-1)^pW_f(\nu)+(-1)^q W_G(\tilde{\mu},\nu')+(-1)^rW_{A_{\tilde{\mu}}}(\nu+\nu')
 \leq (2^{n-1}-2^t)+2^{\frac{n+\lambda}{2}}+( 2^{\frac{n+\lambda}{2}+1}+2^{t+2}),
 \end{align*}
  which is also strictly less than $2^n$ when $t>4$. For any distinct $\tilde{\mu},\tilde{\mu}'\in\mathbb{F}_{2^n}^*$, any $\nu,\nu'\in\mathbb{F}_{2^n}$, and for any $p,q\in\mathbb{F}_2$, by Lemma \ref{Gold-walsh} and Relation \eqref{maxWA}, we have
  \begin{align*}
  W_{A_{\tilde{\mu}}}(\nu)+(-1)^pW_G(\tilde{\mu}',\nu')+(-1)^qW_{A_{\tilde{\mu}+\tilde{\mu}'}}(\nu+\nu')\leq  (2^{\frac{n+\lambda}{2}+2}+2^{t+3})+2^{\frac{n+\lambda}{2}},
  \end{align*}
    which is also strictly less than $2^n$ when $t>4$. Therefore, $C_F$ is minimal by Theorem \ref{thgenericAB}.

    Finally, according to Propositions \ref{propdim} and \ref{propwalshtransform}, Remark \ref{rmkdistance}, Lemma \ref{Gold-walsh}, and Relations \eqref{eqwalsh-f-1E-2} and \eqref{maxWA}, one can obtain that $C_F$ is a $[2^n-1,2n+1,2^{n-2}+2^{t-1}]$ code. This completes the proof.
\end{proof}

Note that the weight distribution of $C_F$ in Theorem \ref{thGold} is not obtained. We invite the interested readers to attack it.

Note also that Lemma \ref{Gold-walsh} holds for any positive integer $n$. Thus, a natural question is that, whether Theorem \ref{thGold} can be generalized to the case $n$ odd. In this case, the function $f$ in Theorem \ref{thGold} cannot be used any more. So one has to find another function $f$ satisfying Conditions $({\bf a})$ and $({\bf b})$ of Construction \ref{ourcon2}, such that the Walsh-Hadamard transform of $ A_{\tilde{\mu}}(x)=f(x)+{\rm Tr}_1^n(\tilde{\mu} G(x))$ can be evaluated for any  $\tilde{\mu}\in\mathbb{F}_{2^n}^*$. It seems very difficult to find such a function, and we leave it for further research.

\begin{remark}
 As three applications of Theorem \ref{thgenericAB}, Theorems \ref{thfirstvioab}, \ref{th-2<m<t} and \ref{thGold} give minimal binary linear codes $C_F$ violating the AB condition for any $2\leq m\leq \frac{n}{2}+1$ and $m=n+1$ when $n$ is even.
  It would be interesting if someone can apply Theorem \ref{thgenericAB} to other cases, including other values $m$ and the case $n$ odd.
\end{remark}

\section{ Concluding remarks}\label{sec:conclusion}

The main contributions of this paper are presented as follows:

 \begin{itemize}
\item A generic construction of minimal linear codes from vectorial Boolean functions (Construction \ref{ourcon}).

\item A necessary and sufficient condition for the linear codes in Construction \ref{ourcon} to be minimal (Theorem \ref{thmainresult}).

\item New three-weight  minimal binary linear codes with complete weight distribution (Corollaries \ref{corsingleplateaued}, \ref{corVB} and \ref{corAB}).

\item A generic construction of minimal linear codes violating the AB condition (Construction \ref{ourcon2}).

\item A necessary and sufficient condition for the linear codes in Construction \ref{ourcon2} to be minimal (Theorem \ref{thgenericAB}).

\item Three infinite families of minimal linear codes violating the AB condition (Theorems \ref{thfirstvioab}, \ref{th-2<m<t} and \ref{thGold}).
\end{itemize}

As remarked in Conclusion of \cite{Ding-2018-IT}, to find an infinite family of minimal linear codes violating the AB condition is a hard problem in general. This paper focus on constructing minimal linear codes from vectorial Boolean functions. To the best of our knowledge, this is the first time that minimal linear codes are constructed from vectorial Boolean functions. Compared with known minimal linear codes, the dimensions of the minimal codes in this paper are in general better.

\section*{Acknowledgments}
This work was supported in part by the National Key Research and Development Program of China under Grant 2019YFB2101703;
by the National Natural Science Foundation of China under Grants 61972258 and U19A2066; by the
Innovation Action Plan of Shanghai Science and Technology under Grants 20222420800 and 20511102200; by the Key Research and Development Program of Guangdong Province  under Grant 2020B0101090001,  and by Scientific Research Fund of Hunan Provincial Education Department under Grant 19B485.

\end{document}